\documentclass[fleqn,usenatbib]{mnras}

\usepackage{newtxtext,newtxmath,hyperref}

\usepackage[T1]{fontenc}
\DeclareRobustCommand{\VAN}[3]{#2}
\let\VANthebibliography\thebibliography
\def\thebibliography{\DeclareRobustCommand{\VAN}[3]{##3}\VANthebibliography}

\usepackage{graphicx}	
\usepackage{amsmath}	
\usepackage{subcaption}
\usepackage{booktabs}
\usepackage{colortbl}
\usepackage[normalem]{ulem}
\usepackage{pdfpages}
\usepackage{multirow}
\usepackage[dvipsnames]{xcolor}

\newcommand*{\kms}{\ensuremath{\text{km}\,\text{s}^{-1}}}
\newcommand*{\msun}{\ensuremath{\text{M}_\odot}}
\newcommand*{\perh}{\ensuremath{h^{-1}}}

\definecolor{Magenta}{rgb}{0.5,1,1}

\newcommand*{\mailto}[1]{\href{mailto:#1}{#1}}
\newcommand*{\http}[1]{\href{http://#1}{#1}}
\newcommand*{\https}[1]{\href{https://#1}{#1}}

\defcitealias{fielder2020}{F20}

\title[The Influence of Subhaloes on Host Halo Properties]{The Influence of Subhaloes on Host Halo Properties}

\author[L.~Mezini et al.]{%
Lorena~Mezini,$^{1,2}$\thanks{E-mail: \mailto{lom31@pitt.edu}}
Catherine E. Fielder,$^{3}$\thanks{E-mail: \mailto{cfielder@arizona.edu}}
Andrew R. Zentner,$^{1,2}$
Yao-Yuan Mao,$^{4}$
Kuan Wang,$^{5, 6}$ and \newauthor
Hao-Yi Wu$^{7}$
\\
$^{1}$Department of Physics and Astronomy, University of Pittsburgh, Pittsburgh, PA 15260, USA\\
$^{2}$Pittsburgh Particle Physics, Astrophysics, and Cosmology Center (PITT PACC), University of Pittsburgh, Pittsburgh, PA 15260, USA\\
$^{3}$Steward Observatory, University of Arizona, Tucson, AZ 85721-0065, USA\\
$^{4}$Department of Physics and Astronomy, University of Utah, Salt Lake City, UT 84112, USA\\
$^{5}$Department of Physics, University of Michigan, Ann Arbor, MI 48109, USA\\
$^{6}$Leinweber Center for Theoretical Physics, University of Michigan, Ann Arbor, MI 48109, USA\\
$^{7}$Department of Physics, Boise State University, Boise, ID 83725, USA \\
\vspace{-3em}
}

\pubyear{2023}
\begin{document}
\label{firstpage}
\pagerange{\pageref{firstpage}--\pageref{lastpage}}
\maketitle

\begin{abstract}
Within the $\Lambda$CDM cosmology, dark matter haloes are composed of both a smooth component and a population of smaller, gravitationally bound subhaloes. These components are often treated as a single halo when properties, such as density profiles, are extracted from simulations. Recent work has shown that density profiles change substantially when subhalo mass is excluded. In this paper, we expand on this result by analysing three specific host halo properties -- concentration ($c_{\rm{NFW}}$), spin ($\lambda_{\rm B}$), and shape ($c/a$), -- when calculated only from the smooth component of the halo. This analysis is performed on both Milky Way-mass haloes and cluster-mass haloes in high-resolution, zoom-in, $N$-body simulations. We find that when subhaloes are excluded the median value of (1) $c_{\rm{NFW}}$ is enhanced by $\approx30\pm11\%$ and $\approx77\pm8.1\%$ for Milky Way-mass ($10^{12.1}\,\msun$) and cluster-mass ($10^{14.8}\,\msun$) haloes respectively, (2) $\lambda_{\rm B}$ is reduced for Milky Way-mass by $\approx11\pm4.9\%$ and cluster-mass haloes by $\approx27\pm3.5\%$. Additionally, with the removal of subhaloes, cluster-mass haloes tend to become more spherical as the ratio of minor-to-major axis, $c/a$, increases by $\approx11\pm3.6\%$, whereas Milky Way-mass haloes remain approximately the same shape with $c/a$ changed by $\approx1.0\pm5.8\%$. Fractional changes of each of these properties depend primarily on the amount of mass in subhaloes and, to a lesser extent, mass accretion history. Our findings demonstrate that the properties of the smooth components of dark matter haloes are biased relative to the total halo mass.

\end{abstract}


\section{Introduction}
\label{section:intro}

Within the standard $\Lambda$CDM cosmology (cold dark matter with a cosmological constant), dark matter haloes form hierarchically. What begins as a peak in the initial density field will eventually collapse under gravity, accumulate matter through mergers with other haloes and form a gravitationally bound, virialized object. Smaller haloes that merge with larger haloes are tidally stripped on their in-fall. Those that are not entirely disrupted by the tidal forces remain self bound and orbit the larger halo (see e.g., \citealt{kauffmann1993,zentner2003,zentner2005a,
bullock2010,Green_2021}). As a result of this hierarchical formation, dark matter haloes are composed of two parts (i) a smooth component, the host halo, composed of disrupted haloes and material from smooth accretion \citep{wang2011} and (ii) subhaloes -- smaller, gravitationally bound clumps within the virial radius of the host.

In this work, we study the influence of subhaloes on measurements of macroscopic halo properties. It is believed that subhaloes can have a spatial and velocity bias with respect to the total dark matter distribution in a halo \citep{diemand2004,zentner2005a}. If this is the case, the properties of a halo including subhaloes will not be representative of the smooth component of the halo. To provide an unbiased view of halo properties, we re-calculate host halo concentration, shape, and spin properties after having removed all mass belonging to subhaloes. This is done by following the technique first introduced in \citet{wu2013} and also employed in \citet[hereafter \citetalias{fielder2020}]{fielder2020}.

Many theoretical studies of haloes are performed with the use of $N$-body \citep[e.g.,][]{1991frenkwhite,kauffmann1993,cole1994,delucia2004,gao2004,zentner2005b,guo2013,klypin2016,ludlow2019}, semi-analytic \citep[e.g.,][]{somerville_primack_1999,taylor2002,zentner2005a,jiang2014,jiang2022}, and hydro-dynamical \citep{Tissera1998,zheng2005,ludlow2020} simulations. 
Within simulation analyses, it is common practice to calculate halo profiles by spherically averaging all the mass associated with a halo, including substructure. However, it is not necessarily the case that substructure traces the smooth component associated with the host halo \citep{diemand2004,zentner2005a,zentner2005b,Nagai_2005}. In fact, \citetalias{fielder2020} has shown the density profile of the smooth component of a halo alone differs significantly from the density profile computed by including the mass of both the smooth component and the subhaloes.  

The difference between the distribution of smooth mass within a halo and the distribution of subhaloes may pose challenges in comparisons between theoretical predictions and observations. While simulation analyses calculate halo profiles through spherical averaging, observational studies will assume a profile model that is smooth and then fit with parameters taken from data \citep[e.g.,][]{moller2002,limousin2006} resulting in two potentially different profiles. Furthermore, some observational gravitational lensing analyses assign subhaloes their own density profiles on top of those of their hosts \citep[e.g.,][]{Newman2013,nirenberg2017,despali2018,gilman2019} which results in double counting of subhaloes. In this follow-up paper to \citetalias{fielder2020}, we aim to alleviate some of this inconsistency by determining the influence of subhaloes on macroscopic halo properties -- concentration, shape, and spin. Previous work, such as \citet{zentner2005a} and \citet{mao2015}, found that subhalo abundance is correlated with host halo concentration, and \citet{fielder2018} found that subhalo abundance is correlated with shape and spin in addition to concentration. The motivation for selecting shape and spin is discussed further below.

Mergers serve as a critical mechanism in the assembly of both galaxies and dark matter haloes. Standard galaxy formation theory suggests that central galaxies form in the potential wells of host haloes, and satellite galaxies in those of subhaloes. Consequently, the formation histories of galaxies are linked to the properties of the haloes in which they form. We take an elliptical galaxy as an illustrative example. Ellipticals are thought to form as the product of merging disk galaxies \citep{Toomre1977, 1982farouki, 1983negroponte, 1993kauffmann, 1998kauffmann}. The merging disk galaxies likely originate from correlated directions \citep{2011libeskind, 2018wangkang}, which influence the resultant elliptical galaxy and dark matter halo causing them to have their principle axes preferentially aligned with large-scale filamentary structure in the cosmic web \citep{zentner2005b, Tempel2015, 2020peng}. Subsequent mergers of satellite galaxies will also preferentially take place along this axis. One can then assume that it is more likely to find satellites along the major axis compared to the minor \citep{zentner2005b, 2006yangxiaohu, 2007faltenbacher, 2008bailin, 2010agustsson, 2021tenneti, 2022gu}. From this example, it is evident that substructure induces a bias in shape that cannot be captured with a smooth density profile alone. Halo shape is one such parameter that is explored in this study.

Understanding the bias induced by substructure on shape is also important for weak lensing observation. Because galaxies are luminous tracers of large-scale structure within the Universe, they serve as a means to study the formation of cosmic structure, the evolution of galaxies, and the nature of dark matter. In the weak lensing regime, galaxy images are distorted by a small amount in an effect called \textit{shear} that results in a systematic alignment of neighboring galaxies that are lensed by the same foreground object. The bias induced in halo shape by the intrinsic alignment of substructure with their host may be correlated with effects that mimic the shear effect of weak lensing \citep{2015troxel, 2021ghosh, 2022zhang}.

The classic theory of disk galaxy formation suggests that the angular momentum of a galaxy is determined by that of its host halo \citep{navarro1991, navarro1993, navarro1997, D_Onghia2006, kaufmann2007}. This concept serves as the basis for using halo spin to approximate galaxy spin in semi-analytic models of galaxy formation. However, some research suggests that this connection is not as direct as initially hypothesized. In simulations, it has been found that there is an average misalignment between the galaxy and halo spin axis of about 30$^{\circ}$ \citep[e.g.,][]{vandenBosch2002a, bett2010} and up to 48.3$^{\circ}$ \citep{Croft2009}. Using observations of nearby star-forming galaxies, \citet{romeo2022} found that baryons contained in the discs and bulges of galaxies retain $\sim$ 80\% of the specific angular momentum of their host halos. However, when broken down, they find that stars have about 40\% less specific angular momentum and atomic gas about 20\% more than the host halo. In order to have reliable and accurate galaxy formation models, we must take into account the degree to which halo spin is not a good proxy for galaxy spin. In this study, we approach this problem by looking at how subhaloes may bias host spin.

The goal of this work is to investigate the effects subhaloes have on halo concentration, shape, and spin as a means to probe the nuanced connection between galaxy and halo evolution and their implications for simulation and observational research. To do so, we measure how these properties would change for a halo if all of its subhaloes were removed, as well as how this change depends on the mass fraction in subhaloes and mass accretion history. In \autoref{Section:Simulations} we discuss the simulations used to perform this analysis. In \autoref{Section:subsremoved} we summarise the subhalo exclusion procedure employed in \citetalias{fielder2020} and describe how we calculate the properties of interest -- concentration, shape, and spin. In \autoref{Section:Analysis} we discuss the measured changes in these properties once subhaloes are removed and their dependence on mass and halo mass accretion history. Finally, conclusions and implications are discussed in \autoref{Section:conclusion}. We include supplemental tables and figures in the appendices.

\section{$N$-body Simulations}
\label{Section:Simulations}
In the following section we describe the simulation data and halo finder used for these analyses and the procedures used to construct our final data sets. The two simulations we discuss are part of the Symphony suite of cosmological zoom-in simulations \citep{nadler2022}.

\subsection{Milky Way and Cluster Mass Zoom-In Simulations}
For this work we utilise two sets of zoom-in cosmological simulations, each representing a different host halo mass range -- Milky Way mass and cluster mass. This is the same set of simulations explored in \citetalias{fielder2020}. These narrow mass ranges allow us to explore subhalo effects that would otherwise be obscured due to the mass trend. By working in two different mass regimes we can ensure consistency of results and test for mass dependence of any observed effects.

\subsubsection{Milky Way Mass haloes}
We used 45 high resolution Milky Way-mass zoom-ins originating from a c125-2048 parent box run with L-GADGET (see \citealt{becker2015}), which were first presented in \citet{mao2015}. These haloes fall within the mass range $M_{\rm VIR} =  10^{12.1 \pm 0.03}\ \mathrm{M}_{\odot}$ and will be referred to in this paper as the \citet{mao2015} Milky Way-mass Zoom-ins, MMMZ. The cosmological parameters for the simulations are $\Omega_{\rm M} = 0.286$, $\Omega_{\Lambda} = 1 - \Omega_{\rm M} = 0.714$, $h=0.7$, mass fluctuation amplitude $\sigma_{8} = 0.82$, and scalar spectral index $n_{\rm s} = 0.96$. They have a particle mass of $m_{\rm p} = 3.0 \times 10^{5}\;\mathrm{M}_{\odot}\,\perh$ and a softening length of 170\,pc\,\perh co-moving, which translates to a lower limit in $V_{\rm max}$ for convergence of approximately 10\,\kms. The resolution limit is taken to be four times the softening length, or 0.68 h$^{-1}$. For more details on these simulations, refer to \citet{mao2015}.

\subsubsection{Cluster Mass haloes}
The cluster-mass halo set contains 96 host haloes within the mass range $10^{14.8\pm0.05}h^{-1}\,\msun$ from the RHAPSODY cluster zoom-ins first presented in \citet{wu2013}. These zoom-ins are high resolution re-simulations of cluster forming regions in one of the Carmen simulations from the LArge Suite of DArk MAtter Simulations \citep{mcbride2009}. For this work, we select the higher resolution re-simulation (RHAPSODY 8K). The Carmen simulation has a volume of 1 $h^{-1}$Gpc and 1120$^3$ particles, whereas, for the same volume, the RHAPSODY 8k has a resolution of 1.0 × 10$^9h^{-1}\,\msun$ (4096$^3$ particles). RHAPSODY 8K has a particle mass of $m_{p} = 1.3\times10^{8}h^{-1}\,\msun$ and a force resolution, as defined above, of 13 $h^{-1}$ kpc. These simulations use a $\Lambda$CDM cosmology with parameters $\Omega_{\rm M} = 0.25$, $\Omega_{\Lambda}$ = 0.75, $h=0.7$, mass fluctuation amplitude $\sigma_{8} = 0.8$, and scalar spectral index $n_{\rm s} = 1.0$. Because halo properties at z=0 depend weakly on cosmological parameters \citep[e.g.,][]{ludlow2014}, the slight variations of this cosmology from the MMMZ cosmology will not result in significant effects when we compare results using the two different halo sets. Further details on how the re-simulations were run can be found in \citet{wu2013}.

\subsection{Halo Identification}
Dark matter haloes in these simulations were identified by the $\tt{ROCKSTAR}$ halo finder \citep{behroozi2013}. In summary, $\tt{ROCKSTAR}$ is a 6D phase-space based finder, which enables robust identification of subhaloes and identifies halo and subhalo relationships with a spherical-overdensity based algorithm. This is advantageous for us because we are interested in studying the properties of hosts separately from their subhaloes; information on both positions and velocities enables us to disentangle subhaloes from their hosts. $\tt{ROCKSTAR}$ produces catalogues of the haloes identified as well as tables of all the particles associated with each halo (i.e. particles that are most bounded to that halo). A more detailed description of $\tt{ROCKSTAR}$ can be found in \citet{behroozi2013}; the source code is also publicly available at \https{bitbucket.org/gfcstanford/rockstar}.

We use the catalogues of dark matter particle positions and velocities that are generated by $\tt{ROCKSTAR}$ (version 0.99.9-RC3+) to analyse host concentration, shape, and spin without the presence of subhaloes, which we refer to as the \textit{smooth} halo component. Using the same definition as \citetalias{fielder2020}, we use the particle catalogues to identify the smooth halo component as mass not associated with any $\tt{ROCKSTAR}$ selected subhalo. It is worth noting that in $\tt{ROCKSTAR}$ a subhalo must have $>50\%$ of its particles bound, meaning that our definition of the smooth component can include loosely bound objects, streams, and caustics -- objects actively undergoing disruption by the host halo.

While there is no explicit way to define the smooth halo component, our working definition depends on the $\tt{ROCKSTAR}$ interpretation of halo identification. The employment of other halo finders and subhalo self-binding criteria could yield quantitatively different results on what objects count as ``substructure.'' However, we expect our results to hold qualitatively due to the overall small subhalo mass fraction relative to the host. Furthermore, studies such as that by \citet{onions2012} show that substructure abundances agree well across halo finders, and \citet{behroozi2013} shows that effects do not manifest until the substructure binding criteria are below 15\%.

Our $\tt{ROCKSTAR}$ based definition of the smooth halo component translates into three different groupings of halo particles.
\begin{itemize}
    \item \textit{subhalo included:} particles associated with the host halo, including particles associated with subhaloes. \\
    \item \textit{subhalo excluded:} particles that are associated with the host halo but \textit{not} associated with any subhaloes identified by $\tt{ROCKSTAR}$. \\
    \item \textit{subhalo only:} particles associated with at least one subhalo, as identified by $\tt{ROCKSTAR}$.
\end{itemize}

By definition, this means that combining the particles in the \textit{subhalo excluded} and \textit{subhalo only} samples yields the full \textit{subhalo included} set of particles. In $\tt{ROCKSTAR}$, all particles in the groups listed above are defined to be within the virial radius of the host halo. Because $\tt{ROCKSTAR}$ uses density criteria to define halo size, removing the mass associated with subhaloes may result in changes to the halo size. As in \citetalias{fielder2020}, we choose to use the virial radius and mass of the host halo listed in the $\tt{ROCKSTAR}$ catalogue throughout all calculations. This maintains consistency between calculations with and without subhaloes and allows us to prescribe modifications that can be applied directly to $\tt{ROCKSTAR}$ catalogues. In \autoref{sec:appendixE}, we include results where virial radii and mass for hosts have been recalculated after subhaloes were removed. For further discussion on the ambiguities in defining these particle subsets we refer the reader to Sections 4.4 and Appendix B of \citetalias{fielder2020}.

\section{Host Halo Properties}
\label{Section:subsremoved}

To study the relationship between host haloes and subhaloes, we compare halo properties derived from the smooth component (subhaloes excluded) to those of the smooth component plus substructure (subhaloes included). The halo properties at the focus of this study are concentration, shape and spin. Throughout the remainder of the paper, we will refer to these as \textit{primary properties} to distinguish from the \textit{secondary properties} further discussed in section \ref{Section:imp_sec_prop}.

\subsection{Computing Primary Halo Properties}
In order to directly determine how subhaloes affect their hosts, we must calculate the host halo properties -- concentration, shape, and spin -- with and without subhaloes present. These calculations are performed directly on the simulation particles. We will use a ``$\dagger$" superscript to denote the properties that were calculated without the presence of subhaloes, e.g., $c_{\text{NFW}}^{\dagger}$ (see below) in contrast to $c_{\text{NFW}}$.

We introduce several cuts to our data set to make sure that all haloes are resolved, in a relaxed state, and do not include any unrealistic features that may have occurred during the simulation process. To ensure that the subhaloes included in our calculations are well resolved, we only include subhaloes with a mass > 10$^{-3} M_{\rm VIR}$ of the host. These thresholds correspond to where the subhalo mass function for haloes in each mass range is a single power law and each subhalo is resolved with at least $\sim 4000$ particles. Additionally, we exclude host haloes that have at least one subhalo with mass greater than 0.2 times its host virial mass to avoid scenarios where the host is not relaxed due to the presence of a large subhalo. Four cluster mass haloes were removed from our set due to this requirement, leaving us with 92. This includes one halo which contains a subhalo more massive than its host -- likely a mislabelling by the halo finder.

Concentration is calculated from a fit to the NFW profile \citep{nfw1996,nfw1997} by maximising the log-likelihood using the Python $\tt{scipy.optimize.minimize\_scalar}$ function with a tolerance of 10$^{-5}$. The NFW profile is one of the most widely used halo density profiles, which is quantified as a measure of halo mass density as a function of distance from the halo centre. The NFW profile is given by
\begin{equation}
\rho_{\rm NFW} = \frac{\rho_{\rm s}}{\frac{r}{r_{\rm s}}(1+\frac{r}{r_{\rm s}})^{2}},
\label{eq:nfw}
\end{equation} 
where $\rho_{\rm s}$ is the characteristic over-density and $r_{\rm s}$ is the scale radius where $d \ln \rho/d \ln r \vert_{r_\text{s}}= -2$. The ratio $R_{\rm VIR}/r_{s}$ is equal to the dimensionless concentration parameter $c_{\rm NFW}$. Our likelihood function is as follows
\begin{equation}
    \mathscr{L}
 = a -  \sum_{i} \textrm{ln} \left ( \frac{x_i}{(1+c x_i)^{2}}\right), 
\end{equation}
where $c$ is concentration and $x_i$ is the radial distance of the $i^\text{th}$ particle from the halo centre, normalised by the virial radius. The normalisation factor, $a$, ensures the NFW distribution behaves as a probability density function (i.e., integrated to 1 from $r=0$ to $R_\text{VIR}$) for any value of $c$. It is defined as
\begin{equation}
    a = \textrm{ln} \left (\frac{1/(1+c) + \ln(1+c) -1}{c^{2}} \right).
\end{equation}

To calculate shape, we adopt the same algorithm as that in $\tt{ROCKSTAR}$ \citep{behroozi2013}, a calculation informed by the findings of \citet{Zemp_2011}. The $\tt{ROCKSTAR}$ algorithm excludes subhaloes when calculating shape. Since the goal of this analysis is to study how shape changes with the exclusion of subhaloes, we \textit{do not} always exclude subhaloes as $\tt{ROCKSTAR}$ does. An essential part of measuring shape is calculating the mass distribution tensor, defined as:
\begin{equation}
M_{ij} = \frac{1}{N}\sum_{N}x_{i,n}x_{j,n},
\end{equation}
where $x_{i,n}$ and $x_{j,n}$ are the coordinates of the $n^{\rm th}$ particle. The eigenvalues of the matrix $M_{ij}$ are equivalent to the squares of the ellipsoid principle axes ($a$, $b$, and $c$, from longest to shortest principle axis). Our measure of shape is the ratio of the shortest to longest axis, $c/a$. In an \autoref{sec:appendixD}, we discuss a variety of similar, but subtly distinct methods for defining the shapes of haloes, including those which use inverse distance squared weighting of particles \citep[e.g.][]{1991dubinski_carlberg,2004kazantzidis+} to mitigate the impact of large substructures on shape measurements. We choose the aforementioned definition of shape because it is the default definition in the widely-used $\tt{ROCKSTAR}$ halo finder and it is the same or similar to the definitions used in the majority of the literature on halo shapes \citep[e.g.][]{1991katz,1991dubinski_carlberg,allgood2006,Kuhlen2007}. An alternative quantification of halo shape is to describe the contours of constant gravitational potential. Potential contours may be of more direct physical relevance for many applications including orbital and galactic dynamics \citep{2007hayashi+}. In addition, the homeoidal theorem implies that iso-potential contours will be both more spherical and vary more smoothly with radius\citep{BT08} 
and that iso-potential shapes are less sensitive to large substructures \citep{2007hayashi+}.

Finally, as our measurement of angular momentum we use the Bullock spin parameter as given in \cite{2001}:
\begin{equation}
    \label{eq:spin}
    \lambda_{\rm B} = \frac{ J_{\rm tot}}
    {\sqrt{2} M_{\rm VIR}R_{\rm VIR}V_{\rm VIR}},
\end{equation}
where the total angular momentum is computed as
\begin{equation}
\Vec{J}_{\rm tot} = m\sum\limits_{i=1}^{n} \overrightarrow{\boldsymbol{r_{i}}} \times \overrightarrow{\boldsymbol{p_{i}}},
\end{equation}
in which $\overrightarrow{r_{i}}$ and $\overrightarrow{p_{i}}$ are the position and momentum of the $i^{\rm th}$ particle and m is particle mass. We have chosen to use the Bullock spin over the traditional spin introduced in \citet{peebles1971} because the Bullock spin does not require knowledge of the total halo energy---a quantity that was not included in the existing halo catalogues.

For each halo, we compute each of the primary halo properties from two particle catalogues. 
The first set of primary halo properties ($c_{\mathrm{NFW}}$, $c/a$, and $\lambda_{\mathrm{B}}$) 
is derived from the {\em subhalo included} set of particles and therefore matches 
the standard definitions for these quantities found in the vast majority of the 
literature on halo properties. The second set of primary halo properties 
(labeled $c_{\mathrm{NFW}}^{\dagger}$, $c/a^{\dagger}$, and $\lambda_{\mathrm{B}}^{\dagger}$), 
is derived from the {\em subhalo excluded} particles. These quantities represent measurements of 
halo properties after excluding all particles associated with subhalos. In computing this 
second set of properties with subhaloes removed, the values of the host virial masses ($M_{\rm VIR}$), 
and host virial radii ($R_{\rm VIR}$) were held fixed, as opposed to being recalculated 
with the reduced set of particles. This choice is the simplest to interpret and likely the 
choice most relevant to physical applications because the differences between the subhalo-included 
and subhalo-excluded properties reflect changes in physical quantities. For example, the 
concentration parameter is defined as the scale radius normalised by the virial radius, 
$c_{\mathrm{NFW}}=R_{\mathrm{VIR}}/r_{\mathrm{s}}$. Therefore, for small changes in 
concentration, the fractional change in concentration and the fractional change in 
scale radius have the same magnitude and it is the change in the scale radius which 
is physically relevant as it sets the scale of the density profile. For those interested, \citetalias{fielder2020} provides an in depth analysis into the changes of the halo density profiles when subhaloes are removed.
This correspondence is even clearer for spin, which is simply a scaled angular 
momentum, so the fractional change in spin is identical to the fractional change 
in angular momentum, which is the physically relevant quantity. Nonetheless, it may 
be interesting to show the degree to which these scaled halo properties change when 
they are recomputed without subhalos {\bf and} the host halo virial radii and masses 
are {\bf also} recomputed from the subhalo-excluded particle set. We include these 
results in \autoref{sec:appendixE} for completeness.

\subsection{Impact of Secondary Properties}
\label{Section:imp_sec_prop}

We are interested in how the aforementioned three primary halo properties (concentration, shape, and spin) change when subhaloes are not included. However, the magnitude of change in these properties can depend on a number of secondary properties. In this paper the secondary properties we explore include (1) the halo mass fraction in subhaloes and (2) halo mass accretion history. 

The mass fraction in subhaloes is the fraction of the total mass of all particles that explicitly belong to subhaloes relative to the total mass of the entire system (host + subhaloes). This is a natural choice of properties to explore, as we anticipate the magnitude of primary halo property changes to correlate with the relative mass fraction in subhaloes.

Mass accretion history serves as a means of quantifying the relaxation state of a halo. Because haloes with more quiescent merger histories are in a more relaxed state, we anticipate that they will have fewer subhaloes and exhibit smaller changes in their primary properties when subhaloes are excluded. Mass accretion is quantified by the fraction of present-day mass the halo has acquired as a function of time. We use the cosmic scale factor of the Universe as our ``time" variable, which is related to redshift by $a=1/(1+z)$, where $z$ is redshift and $a$ spans from $a=0$ at the Big Bang to $a=1$ at present. We examine the scale factors when the halo system has accumulated either 25\%, 50\%, 70\%, or 90\% of its present-day, $z=0$ mass. The value of the scale factor when a halo first acquires a particular fraction of mass is often used as a definition for halo ``formation time" or halo ``formation scale factor." We follow the common convention and use these terms interchangeably.

We are interested in seeing if the change in our primary properties -- concentration, shape, and spin -- when subhaloes are removed depends on the mass fraction in subhaloes or mass accretion history. To do so, we calculate the fractional change in the primary properties as a function of the two secondary properties. Because we cannot assume that our data will follow a particular functional form, we chose to calculate the Spearman rank-order coefficient in order to determine if there is some correlation. This was done using $\tt{scipy.stats.spearmanr}$.

\section{Results}
\label{Section:Analysis}

In \citetalias{fielder2020}, it is demonstrated that the removal of subhaloes results in significant changes to the host halo density profile. In this section we explore the subhalo effect on host halo primary properties. We then quantify these effects by the host halo secondary properties. 

\subsection{Impact of Subhaloes on Host Halo Properties}
\label{subsection:imp_primary}

\begin{figure*}
\centering
\begin{subfigure}{\textwidth}
    \includegraphics[width=\textwidth]{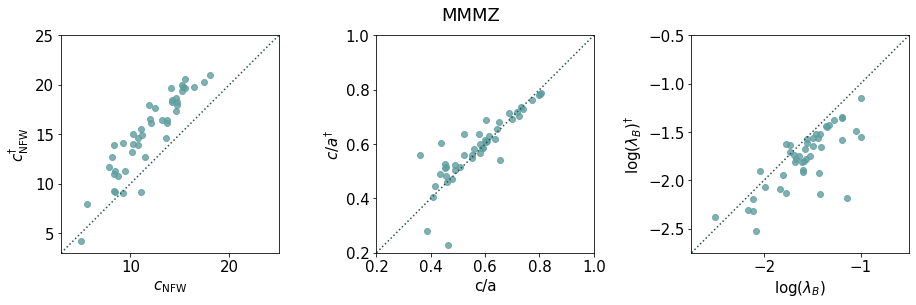}
    \caption{}
    \label{subfigure:mmmz_scatter}
\end{subfigure}
\begin{subfigure}{\textwidth}
    \includegraphics[width=\textwidth]{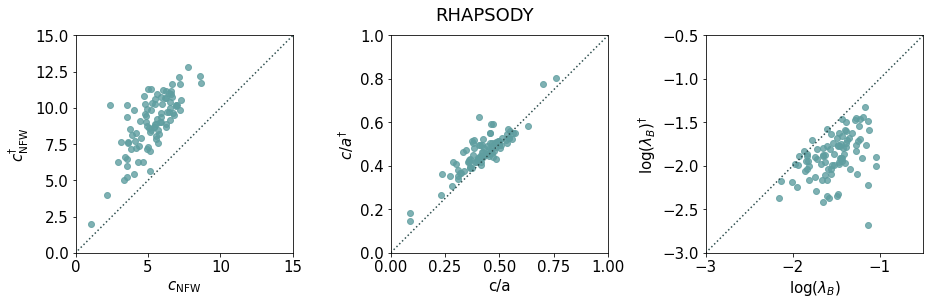}
    \caption{}
    \label{subfigure:rhap_scatter}
\end{subfigure}
\caption{Scatter plots of halo concentration, shape, and spin with and without the presence of subhaloes for the Milky Way-mass (a) and cluster-mass (b) haloes. Halo property values including subhaloes are plotted along the x-axis and those excluding subhaloes along the y-axis. Each teal circle represents a halo in our simulations. Haloes that fall along the dotted diagonal line did not experience any change in a given property when its subhaloes were removed. We show that in most cases haloes behave similarly when their subhaloes are removed. \textbf{(a):} In the left panel, concentration increases for all but three haloes when subhaloes are not included. In the middle panel, haloes are scattered around the dotted line with no noteable shift in one direction. Finally, in the right panel, there is a decrease in spin for all but five haloes when subhaloes are removed. \textbf{(b)}: The lower row of panels depicts our results for the cluster-mass RHAPSODY haloes. In the left panel, concentration increases for all haloes when subhaloes are not included. In the middle panel, there is a lot of scatter however the majority of haloes fall above the dotted line indicating a shift in shape towards unity. In the right panel, there is a decrease in spin for most haloes when subhaloes are removed.}
\label{figure:prop_change_scatter_plots}
\end{figure*}

First, we examine the way in which removal of the mass associated with subhaloes 
alters the primary properties of host dark matter haloes on a halo-by-halo basis. 
Figure~\ref{figure:prop_change_scatter_plots} shows scatter plots of each of our primary 
properties computed in the standard manner, compared with the values of these primary 
halo properties computed after removal of substructure (labeled with superscript ``$\dagger$"). 
It is clear that removing substructure and focusing on the mass associated with the 
smooth component of the host halo results in increased concentrations (left panels), 
moderately more spherical shapes (middle panels), and reduced spins (right panels) for 
both the Milky Way-mass MMMZ haloes (top row) and the cluster-mass RHAPSODY haloes (bottom row).

\begin{figure*}
\centering
\begin{subfigure}{\textwidth}
    \includegraphics[width=\textwidth]{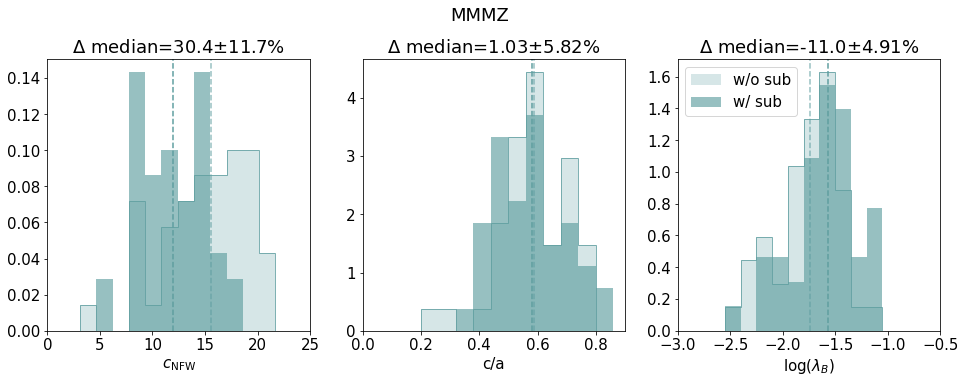}
    \caption{}
    \label{subfigure:mmmz_hist}
\end{subfigure}
\begin{subfigure}{\textwidth}
    \includegraphics[width=\textwidth]{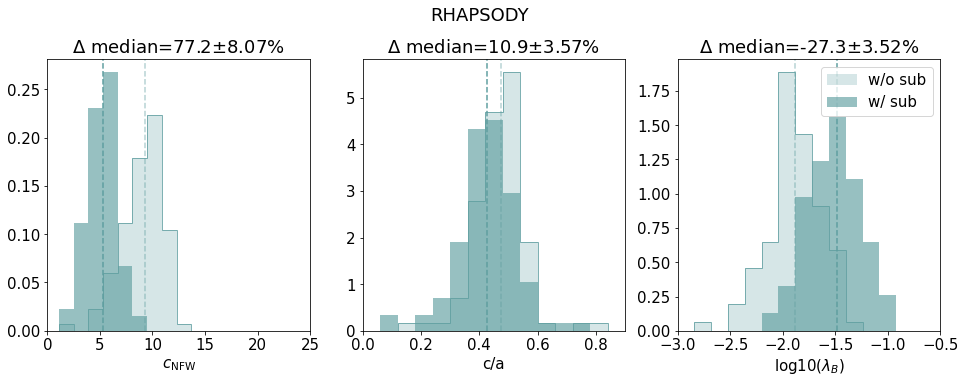}
    \caption{}
    \label{subfigure:rhap_hist}
\end{subfigure}
\caption{Histograms of halo concentration, shape, and spin with and without the presence of subhaloes for the Milky Way-mass (a) and cluster mass (b) haloes. The $\Delta$ in the titles represents the difference in medians between the two distributions. The dark teal histograms represent the halo properties including subhaloes (w/ sub) and the light teal excluding subhaloes (w/o sub). The dashed vertical lines indicate the median value of each histogram. \textbf{(a):} In the left panel, concentration increases when subhaloes are not included as the density profile of the halo becomes more centrally dense. In the middle panel, there is no notable shift in the distribution of minor-to-major axis ratios. Finally, in the right panel, there is a decrease in spin when subhaloes are removed, which is expected because the angular momentum imparted onto host haloes through their subhaloes will be driven down when subhaloes are removed. \textbf{(b)}: The lower row of panels depicts our results for the cluster-mass RHAPSODY haloes. In the left panel, concentration increases when subhaloes are not included as the density profile of the halo becomes more centrally dense. In the middle panel, there is a shift in the shape distribution towards unity, indicating that haloes are becoming more spherical. In the right panel, there is a decrease in spin when subhaloes are removed, which is expected because the angular momentum imparted onto host haloes through their subhaloes will be driven down when subhaloes are removed.}
\label{figure:histograms}
\end{figure*}

Next, we examine the distributions of host halo primary properties for both the Milky Way-mass (MMMZ) and cluster-mass (RHAPSODY) haloes, as shown in \autoref{subfigure:mmmz_hist} and \autoref{subfigure:rhap_hist}. Both figures have three histogram plots, one for concentration ($c_{\rm{NFW}}$; left), shape (c/a; middle), and spin ($\lambda_{\rm{Bullock}}$; right). For each property we show the distribution of values including subhaloes (dark teal) and excluding subhaloes (light teal). Vertical dashed lines indicate the median value of each histogram, which are given in \autoref{table:mwm_medians} and \autoref{table:rhap_medians} for MMMZ and RHAPSODY haloes respectively. Alternating rows in these tables give the values with subhaloes included and excluded. In the rightmost column we quantify the dispersion among data points by using the interquartile range (IQR). To assess whether there is a significant difference between subhalo included and excluded distributions, we perform a two-sided Mann-Whitney U test using $\tt{scipy.stats.mannwhitneyu}$. Results from this test are given in the following text.

\begin{figure*}
\centering
\begin{subfigure}{0.45\textwidth}
    \includegraphics[width=\textwidth]{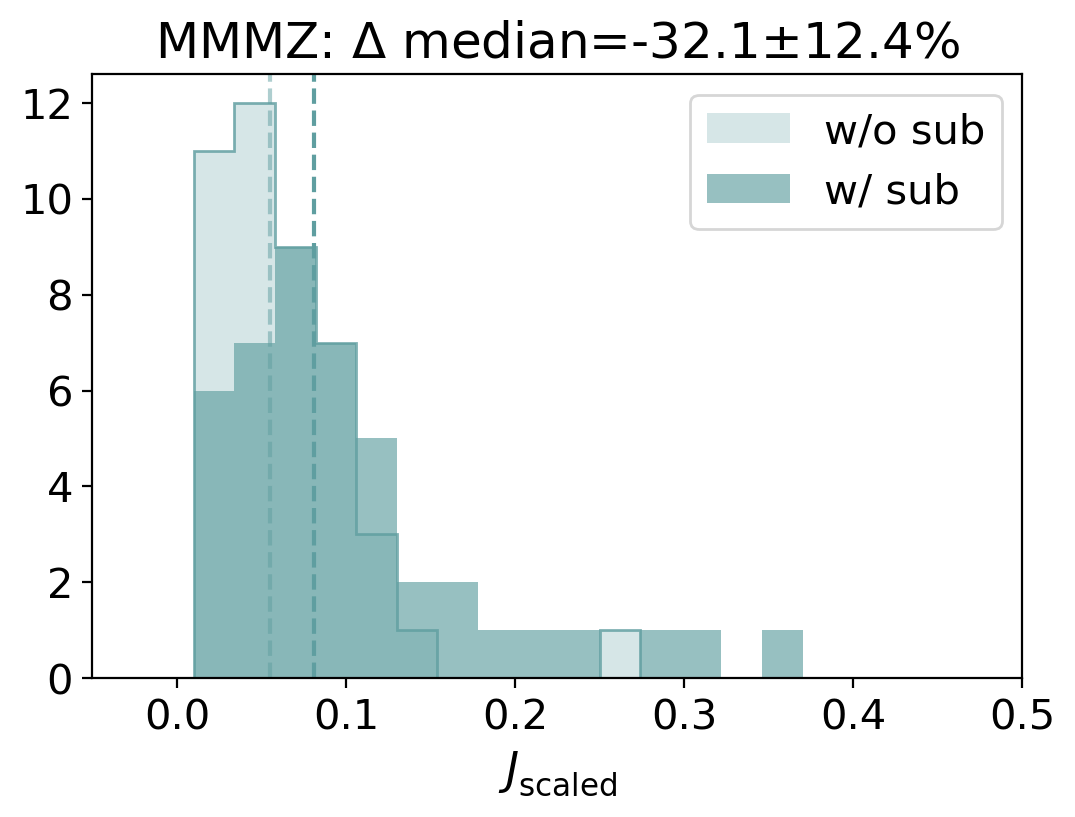}
    \caption{}
    \label{subfigure:mmmz_ang_mom_hist}
\end{subfigure}
\begin{subfigure}{0.45\textwidth}
    \includegraphics[width=\textwidth]{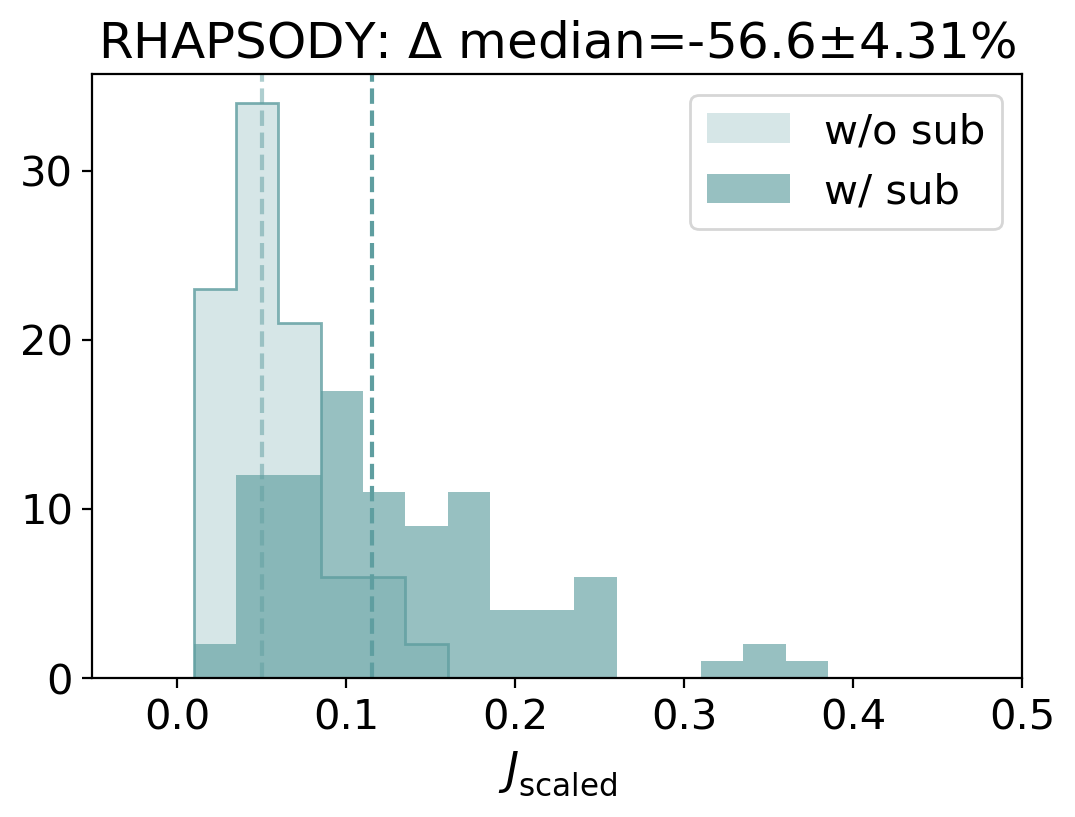}
    \caption{}
    \label{subfigure:rhap_ang_mom_hist}
\end{subfigure}
\caption{Histograms of total angular momentum scaled by $\sqrt{2GM^3R}$ where G is the gravitational constant and M and R are a fiducial mass and radius. The dark teal histograms represent the halo properties including subhaloes (w/ sub) and the light teal excluding subhaloes (w/o sub). The dashed vertical lines indicate the median value of each histogram. In both cases, angular momentum decreases upon removal of subhaloes. This is expected because the angular momentum imparted onto host haloes through their subhaloes will be driven down when subhaloes are removed.}
\label{figure:ang_mom_histograms}
\end{figure*}

\begin{table*}
\centering
\large
\setlength{\tabcolsep}{23pt}
\renewcommand{\arraystretch}{1.5}
\begin{tabular}{ ccc }

 \multicolumn{3}{c}{\textsc{MMMZ Halo Property Median Values}}\\

 \rowcolor{purple!25} \textsc{Property} & \textsc{Median}
 &\textsc{Scatter}\\

 \rowcolor{purple!5} $c_{\rm NFW}$ & 11.9 & 5.34 \\
 \rowcolor{purple!5}$c^{\dagger}_{\rm NFW}$ & 15.5 & 5.58 \\
 
 $c/a$ & 0.581 & 0.187\\
 $c/a ^{\dagger}$ & 0.587 & 0.165\\
 
 \rowcolor{purple!5} $\rm log_{\rm 10}(\lambda_{\rm B})$ & -1.56 & 0.324\\
 \rowcolor{purple!5}$\rm log_{\rm 10}(\lambda_{\rm B})^{\dagger}$ & -1.74 & 0.368\\

 $J_{scaled}$ & 8.13$\times$ 10$^-2$ & 6.83$\times$ 10$^-2$ \\
 $J_{scaled}^{\dagger}$ & 5.52$\times$ 10$^-2$ & 4.99$\times$ 10$^-2$\\
 
\end{tabular}
\caption{Median values of halo concentration, shape, and spin distributions shown in \autoref{subfigure:mmmz_hist} and scaled angular momentum as shown in \autoref{subfigure:mmmz_ang_mom_hist}. Alternating rows give the values with subhaloes included and excluded. The rightmost column gives the scatter calculated via the interquartile range (IQR). Concentration increases when subhaloes are not included as the density profile of the halo becomes more centrally dense. Shape remains approximately the same with a small shift towards more spherical. Finally, there is a decrease in both spin and scaled angular momentum when subhaloes are removed, which is expected because the angular momentum imparted onto host haloes through their subhaloes will be driven down when subhaloes are removed. Scatter decreases when subhaloes are excluded for shape and angular momentum.}
\label{table:mwm_medians}
\end{table*}

\begin{table*}
\centering
\large
\setlength{\tabcolsep}{23pt}
\renewcommand{\arraystretch}{1.5}
\begin{tabular}{ ccc }

 \multicolumn{3}{c}{\textsc{RHAPSODY Halo Property Median Values}}\\
 \rowcolor{purple!25} \textsc{Property} & \textsc{Median}
 &\textsc{Scatter}\\
 
 \rowcolor{purple!5} $c_{\rm NFW}$ & 5.41 & 1.91 \\
 
 \rowcolor{purple!5}$c^{\dagger}_{\rm NFW}$ & 9.21 & 2.46 \\
 
 $c/a$ & 0.425 & 0.107\\
 
 $c/a ^{\dagger}$ & 0.466 & 0.087\\
 
 \rowcolor{purple!5} $\rm log_{\rm 10}(\lambda_{\rm B})$ & -3.44 & 0.65\\
 
 \rowcolor{purple!5}$\rm log_{\rm 10}(\lambda_{\rm B})^{\dagger}$ & -4.34 & 0.70\\

 $J_{scaled}$ & 0.115 & 9.25$\times$10$^{-2}$\\
 $J_{scaled}^{\dagger}$ & 5.01$\times$10$^{-2}$ & 4.21$\times$10$^{-2}$ \\

\end{tabular}
\caption{The same as \autoref{table:mwm_medians} but for the RHAPSODY halo distributions shown in \autoref{subfigure:rhap_hist} and \autoref{subfigure:rhap_ang_mom_hist}. Concentration increases when subhaloes are not included as the density profile of the halo becomes more centrally dense. Shape shifts towards more spherical. Finally, there is a decrease in spin and scaled angular momentum when subhaloes are removed, which is expected because the angular momentum imparted onto host haloes through their subhaloes will be driven down when subhaloes are removed. Scatter decreases for shape and scaled angular momentum when subhaloes are excluded.}
\label{table:rhap_medians}
\end{table*}


Consider first the shifts in concentration with the removal of subhaloes. The concentration shifts are shown in the {\em left} panels of \autoref{figure:histograms}. Notice that for both the MMMZ and RHAPSODY haloes, concentration systematically increases with the removal of subhaloes. Performing a U test shows an increase in concentration with a $p$-value of $\approx 1 \times$10$^{-4}$ for MMMZ and $\approx 2.4 \times$10$^{-24}$ for RHAPSODY, where the $p$-values here are defined to be the probabilities that the two samples are drawn from a common underlying distribution. We take $p$-values less than 0.05 to be significant. These U test results indicate a significant increase in concentration for both halo groups. In the case of the MMMZ haloes (\autoref{subfigure:mmmz_hist}), the shift is relatively subtle. However, after the removal of subhaloes, the distribution of concentrations peaks at higher concentrations and has a tail out to smaller concentrations. The 
change in concentration is more apparent on a halo-by-halo basis 
in the scatter plot of Fig.~\ref{figure:prop_change_scatter_plots}. 
The shift toward higher concentrations upon subhalo removal is more pronounced for the RHAPSODY halo sample (\autoref{subfigure:rhap_hist}). Additionally, the scatter in concentration after subhaloes are removed is smaller in the more massive RHAPSODY haloes, which is consistent with \citet{neto2007} and \citet{duffy2008}. Our results are in agreement with findings from \citetalias{fielder2020} where concentrations were derived via non-linear least squares fitting. 

The shifts in concentration upon subhalo removal that we measure are consistent with expectations. First, subhaloes have long been known to be biased tracers of the overall mass distribution within dark matter haloes. Subhaloes tend to inhabit the outskirts of haloes, so removing subhaloes preferentially removes mass from the outer halo \citep{zentner2005b, Nagai_2005}. This has the effect of making the halo profile more centrally peaked. Second, high-mass haloes have long been known to have a higher proportion of their overall masses bound up in massive subhaloes. Therefore, removing subhaloes results in removing a larger fraction of mass from high-mass, cluster-sized haloes than from comparably lower-mass haloes.

The results for shape (quantified as the minor-to-major axis ratio $c/a$) are given in the {\em middle} panels of \autoref{figure:histograms}. For RHAPSODY haloes, the axis ratio shifts towards greater values, indicating that the haloes become more spherical when subhaloes are removed. This is shown by a modest increase in the median ratio by $\sim$9\% and a vanishing tail in the distribution at smaller values. MMMZ haloes, on the other hand, remain approximately the same shape. We calculate the U statistic for both mass groups and find that RHAPSODY haloes show a significant increase towards more spherical shape with and $p$-value  1.3$\times$10$^{-3}$ and MMMZ haloes show no significant change in shape with $p$-value 0.41.We also computed the angular difference in the major-axis vector including and excluding subhaloes and find that the orientation remains approximately unchanged for most haloes when subhaloes are removed. 

The RHAPSODY results match our expectations for similar reasons to the concentration changes. Subhaloes tend to have more elliptical orbits due to the preferred halo merger direction being along filaments. Because subhaloes are located away from the host centre, this causes the entire halo system to have an overall elliptical shape. With this in mind, it should be expected that the halo system will become more spherical when the components on the outskirts with highly elliptical orbits are removed. There are several reasons that could explain why we do not detect a significant change in shapes for MMMZ haloes. As mentioned above in the discussion on concentration, Milky Way mass haloes generally contain less mass in massive subhaloes than cluster mass haloes. 
Furthermore, Milky Way mass haloes are less likely to have undergone recent mergers, and so are less likely to have subhaloes on highly elliptical orbits.

Results for spin are given in the {\em right} panels of \autoref{figure:histograms}. The Bullock spin decreases for both RHAPSODY and MMMZ haloes, with their median spin value reduced by $\sim$26\% and $\sim$11\% respectively. A calculation of the U statistic shows a significant decrease in spin for both halo groups with $p$-values of 5.94$\times$10$^{-16}$ for RHAPSODY and 1.16$\times$10$^{-3}$ for MMMZ. This change is greater for cluster-mass haloes than Milky Way mass haloes, which makes sense for a number of reasons. Cluster-mass haloes contain a large fraction of their masses in subhaloes. Furthermore, more massive subhaloes are preferentially found at large halocentric distances, so they contribute a significant amount of angular momentum. Host halo formation time may also be a contributing factor. As a result of hierarchical halo formation, Milky Way-mass haloes tend to form at earlier times and are more relaxed than cluster-mass haloes. Hence, Milky Way-mass system subhaloes are more likely to have undergone dynamical friction and lost angular momentum to their hosts compared to the recently merged subhaloes of cluster mass haloes. We also compute the angular difference in the spin vector including and excluding subhaloes and find that the orientation remains approximately unchanged for most haloes when subhaloes are removed.

Angular momenta represented by the spin parameter $\lambda_{\mathrm{B}}$ are scaled by a 
halo mass-dependent unit. For completeness, we also compute the decline in angular momentum. To 
render angular momentum values to be similar in magnitude to common values of the 
spin parameter, we normalise by a fiducial angular momentum $J_{\mathrm{fid}} = \sqrt{2}M_{\mathrm{fid}}V_{\mathrm{fid}}R_{\mathrm{fid}}$, where $R_{\mathrm{fid}}$ is a fiducial 
halo virial radius, $M_{\mathrm{fid}}$ is given by 
\begin{equation}
    M_{\rm fid} = \frac{4}{3}\pi R_{\mathrm{fid}}^3 \Delta_{\rm VIR}\rho_{\rm crit},
\end{equation}
and $V_{\mathrm{fid}} = \sqrt{G M_{\mathrm{fid}}/R_{\mathrm{fid}}}$. 
In this way, the angular momenta of all haloes in each of our samples are given in a 
common set of units. For the cluster-mass haloes of RHAPSODY, we choose 
$R_{\mathrm{fid}} = 1.7\, \mathrm{Mpc}$ (corresponding to a fiducial virial 
mass of $M_{\mathrm{fid}} = 14.42 \times 10^{14} \msun$) while for the Milky Way-mass MMMZ haloes 
we choose $R_{\mathrm{fid}} = 0.2\, \mathrm{Mpc}$ (corresponding to 
$M_{\mathrm{fid}} = 11.63 \times 10^{11} \msun$). In \autoref{figure:ang_mom_histograms}, 
we show histograms of the scaled angular momentum $J_{\mathrm{scaled}} = J_{\mathrm{tot}}/J_{\mathrm{fid}}$ computed both with subhalo particles and 
with subhaloes excluded. 
For both Milky Way-mass and cluster-mass haloes, the scaled angular momentum decreases with the removal of subhalos. Scaled angular momentum is $\sim$56\% less for RHAPSODY and $\sim$32\% less for MMMZ. The Mann-Whitney U statistic indicates a significant decrease in angular momentum with $p$-values of 6.53$\times 10^{-3}$ and 6.51$\times 10^{-16}$ for MMMZ and RHAPSODY, respectively.

\subsection{Correlation with Secondary Properties}
\label{subsection:res_sec_prop}

In this subsection we explore to what extent the changes in host halo primary properties (halo concentration, shape, and spin) are affected by our secondary properties of interest (mass fraction in subhaloes and formation time). Specifically, we calculate the fractional change of our primary properties, as a function of the two secondary properties and determine the Spearman rank-order correlation coefficient between the two. Fractional change is defined as the difference between a property value with and without subhalos, relative to the property value with subhaloes. For example, for concentration this would be
\begin{equation}
    \frac{\Delta c_{\rm{NFW}}}{c_{\rm{NFW}}}  = \frac{c_{\rm{NFW}} - c^{\dagger}_{\rm{NFW}}}{c_{\rm{NFW}}}.
    \label{eq:frac_change}
\end{equation}
\subsubsection{Exploring Secondary Properties}

Results for the correlation of fractional change in primary halo properties with mass fraction in subhaloes are given in \autoref{table:mwm_mass_frac} for Milky Way-mass haloes and in \autoref{table:rhap_mass_frac} for cluster-mass haloes. The middle column of the tables gives the Spearman correlation coefficient and in the right column is the $p$-value corresponding to our null hypothesis that there is no correlation. Both correlation and $p$-value are computed using the $\tt{scipy.stats.spearmanr}$ function. We chose this statistic since it does not require assuming a particular functional form for our data. We consider a measurement with $p$-value < 0.05 to be significant.

In the top row of \autoref{table:mwm_mass_frac} and \autoref{table:rhap_mass_frac}, we see statistically significant positive correlations between the mass fraction in subhaloes and the fractional change in $c_{\rm NFW}$ for both Milky Way- and cluster-mass haloes. As discussed in \autoref{subsection:imp_primary}, subhaloes occupy orbits at the outskirts of haloes. As a result, we expect halo concentration to increase with the exclusion of subhaloes as the mass distribution becomes more centrally peaked, in agreement with \citetalias{fielder2020}. Here we show that not only does concentration increase when subhaloes are removed, but, not surprisingly, the increase in concentration is greater for host haloes that have a higher fraction of mass bound in subhaloes.

The second rows of \autoref{table:mwm_mass_frac} and \autoref{table:rhap_mass_frac} show that haloes with a larger mass fraction in subhaloes tend to become more spherical once subhaloes are excluded. This fits in with our understanding that subhaloes occupy more elliptical orbits \citep{Gill2004, diemand2007, klimentowski2010, Elahi2018} and as these objects are removed, host halo shapes become less ellipsoidal. Because Milky Way-mass haloes contain less mass in subhaloes compared to cluster-mass haloes, this likely accounts for why the overall distribution of shapes shows very little change after subhaloes are removed in \autoref{subfigure:mmmz_hist}. For example, our MMMZ haloes have an average subhalo mass fraction of $\sim$ 6\% compared to $\sim$ 11\% for RHAPSODY haloes. 

The bottom row of \autoref{table:mwm_mass_frac} and \autoref{table:rhap_mass_frac} show that the decrease in spin for Milky Way-mass and cluster-mass haloes depends on the amount of mass removed. This result is consistent with the discussion presented in \autoref{subsection:imp_primary} -- haloes lose the angular momentum imparted by subhaloes when the subhaloes are removed. 

\begin{table*}
\centering
\large
\setlength{\tabcolsep}{23pt}
\renewcommand{\arraystretch}{1.5}
\begin{tabular}{ ccc }

 \multicolumn{3}{c}{\textsc{MMMZ Halo Property Correlation with Subhalo Mass}}\\
 
 \rowcolor{purple!25} \textsc{Property} & \textsc{Coefficient}
 &\textsc{$p$-value}\\
 
 \rowcolor{purple!5} $c_{\rm NFW}$ & 0.416 & \textbf{4.47}$\times$ \textbf{10$^{-3}$} \\

 $c/a$ & 0.429 & \textbf{3.29}$\times$ \textbf{10$^{-3}$}\\

 \rowcolor{purple!5} $\rm log_{\rm 10}(\lambda_{\rm B})$ & 0.475 & \textbf{9.67}$\times$ \textbf{10$^{-4}$}\\

\end{tabular}
\caption{Measure of fractional property change as a function of mass fraction in subhaloes for Milky Way-mass haloes. The middle column gives the Spearman rank order coefficient of the correlation. The right-most column gives the $p$-value corresponding to the null hypothesis that there is no correlation. For each property, there is a significant positive relationship with the amount of mass in subhaloes that is removed, which is indicated also by the $p$-values given in bold font.}
\label{table:mwm_mass_frac}
\end{table*}

\begin{table*}
\centering
\large
\setlength{\tabcolsep}{23pt}
\renewcommand{\arraystretch}{1.5}
\begin{tabular}{ ccc }

 \multicolumn{3}{c}{\textsc{RHAPSODY Halo Property Correlation with Subhalo Mass}}\\
 \rowcolor{purple!25} \textsc{Property} &\textsc{Coefficient} &\textsc{$p$-value}\\
 
 \rowcolor{purple!5} $c_{\rm NFW}$ & 0.466 & \textbf{2.84} $\times$ \textbf{10$^{-6}$}\\
 
 $c/a$ & 0.449 & \textbf{7.26}$\times$ \textbf{10$^{-6}$} \\

 \rowcolor{purple!5} $\rm log_{\rm 10}(\lambda_{\rm B})$ & 0.416 & \textbf{3.79}$\times$ \textbf{10$^{-5}$}\\

\end{tabular}
\caption{Same as \autoref{table:mwm_mass_frac} but for cluster-mass haloes. For each property there is a significant positive relationship with the amount of mass in subhaloes that is removed.}
\label{table:rhap_mass_frac}
\end{table*}

The Spearman rank-order correlation coefficients for mass accretion history are given in \autoref{table:mwm_assembly} and \autoref{table:rhap_assembly} for Milky Way- and cluster-mass haloes respectively. As a proxy for mass accretion history, we use the scale factor of the Universe at various mass accretion percentages. In particular, we use the scale factor when the host halo first accumulated 25\%, 50\%, 70\%, or 90\% of its final, $z=0$ mass. We designate these as $a_{25}$, $a_{50}$, $a_{70}$, and $a_{90}$ respectively. 

For the cluster-mass RHAPSODY haloes, there is no significant relationship between concentration change and any of the formation time proxies. In the case of the Milky Way-mass MMMZ host haloes, we detect a marginally significant positive correlation between the change in concentration and only one of our formation time proxies ($a_{70}$). \cite{wang2020} showed that there is scatter in the relationship between halo formation time and concentrations due to recent mergers which may dramatically alter halo concentration over a dynamical time period. Our difficulty to detect an underlying correlation may stem from this same effect of scatter induced by recent mergers.

For shape, we are not able to detect any significant relationship between the change in shape and the mass accretion history in RHAPSODY haloes or MMMZ haloes. In conjunction with the results for subhalo mass fraction in \autoref{subsection:imp_primary}, it appears that halo shape is influenced more by the amount of mass in subhaloes rather than when those subhaloes were accreted.

There is a significant positive trend between change in spin and formation time at the 70\% threshold for MMMZ haloes. This fits in with our understanding of halo evolution. Angular momentum will generally increase over time as more subhaloes are accreted and impart their angular momentum. For RHAPSODY, there are no significant trends between formation time and spin.

We also use the scale factor at which a halo had its last major merger rather than a specific mass threshold, where we define a major merger between a halo of mass $M$ with another halo of mass $\geq$ 0.3$M$. This choice was physically motivated by the fact that major mergers correspond to a significant and sudden increase in mass. However, there was no significant correlation with the fractional change in halo properties with this metric.


\subsubsection{Modeling Subhalo Exclusion}
\label{subsubsection:corrections}

We want to provide readers with the option to modify their data to explore subhalo effects in their own work. To this end, we use our analysis of the influence of subhalo mass fraction to derive linear fits to the changes in halo properties as a function of mass fraction. The results of the linear fits can be used to adjust halo property values that were calculated including subhaloes to those excluding subhaloes. In the cases where there is a correlation between the change in a halo property with mass fraction, we performed a linear regression using the scikit learn Huber Regressor \citep{scikit-learn}. Because this method is robust to outliers, it is ideal for our small data samples which contain significant scatter. As an additional means to reduce the influence of outliers, we binned our haloes by subhalo mass fraction and performed the fit on the median value per bin. Fit parameters are given in \autoref{table:mwm_mass_frac_mod} and \autoref{table:rhap_mass_frac_mod}, and scatter plots with the best fit line are given in \autoref{sec:appendixB}. Although we have taken steps to make our analysis robust to outliers, it is still possible that the scatter in our data may introduce some bias in our modifying equations.

The first column of each table indicates the halo property of interest and the second and third columns give the slope and intercept from the fit with their errors calculated via bootstrap in parenthesis. The fourth column gives the coefficient of determination, $R^{2}$, of these fits. We expect that as the subhalo mass fraction approaches zero, the fractional change in properties should also approach zero. However, the intercepts of our fits are non-zero values. In addition, the linear behaviour between subhalo mass fraction and property change must break down at small mass fractions, in which case a different fitting approach must be taken. Because we do not have data for very small subhalo mass fractions, we do not anticipate these issues being a problem for our analysis.

The sixth column shows, for each property, the difference between the median values of the subhalo-excluded distribution and the subhalo-included distribution which has been corrected using the linear fits we derived. The differences in medians are substantially smaller than the ones given in \autoref{figure:histograms} for almost all properties, except Milky Way-mass halo shape. This is not unexpected as there was negligible change in the median shape as evidenced in \autoref{subfigure:mmmz_hist}.  When both performance measures (RMSE and median differences) are taken into account, these results indicate that our modifications are effective.

\begin{table*}
\centering
\large
\setlength{\tabcolsep}{15pt}
\renewcommand{\arraystretch}{1.25}
\begin{tabular}{ cccccc }

 \multicolumn{6}{c}{\textsc{MMMZ Halo Property Modification}}\\
 
 \rowcolor{purple!25} \textsc{Property} & \textsc{Slope} &\textsc{Intercept}
 &\textsc{$R^{2}$} & \textsc{RMSE} & \textsc{$\Delta$Median}\\

 \rowcolor{purple!5} $c_{\rm NFW}$ & 1.20 (0.640) & 0.231 (3.17$\times$10$^{-2}$) & 0.792 & 1.55 & -2.25\%\\

 $c/a$ & 0.690 (0.275) & -1.36$\times$10$^{-2}$ (9.55$\times$10$^{-2}$) & 0.809 & 6.99$\times$10$^{-2}$ & 0.508\%\\

 \rowcolor{purple!5} $\rm log_{\rm 10}(\lambda_{\rm B})$ & 0.259 (0.749) & 6.31$\times$10$^{-2}$ (2.32$\times$10$^{-2}$) & 0.211 & 0.227 & 2.92\%\\

\end{tabular}
\caption{Modification parameters for each halo property for MMMZ haloes derived from fits to the fractional property change as a function of subhalo mass fraction. The first column shows the halo property of interest and the second and third columns give the slope and intercept from the Huber's T linear regressions with their errors in parenthesis. The fourth column gives the coefficient of determination, \textsc{$R^{2}$}, of the fits. We use the results from the fits to correct property measurements calculated including subhaloes. The RMSE from comparing the modified property values to the true values excluding subhaloes are given in the fifth column. The sixth column shows the percent change in median values from the true and fit-modified distribution. Except for shape, these differences are substantially smaller than the ones given in \autoref{subfigure:mmmz_hist}.}
\label{table:mwm_mass_frac_mod}
\end{table*}

\begin{table*}
\centering
\large
\setlength{\tabcolsep}{15pt}
\renewcommand{\arraystretch}{1.25}
\begin{tabular}{ cccccc }

 \multicolumn{6}{c}{\textsc{RHAPSODY Halo Property Modification}}\\
 \rowcolor{purple!25} \textsc{Property} & \textsc{Slope} &\textsc{Intercept} &\textsc{$R^{2}$} & \textsc{RMSE} & \textsc{$\Delta$Median}\\

 \rowcolor{purple!5} $c_{\rm NFW}$ & 2.39 (0.592) & 0.289 (6.66$\times$10$^{-2}$) & 0.711 &  1.35 & 4.03\%\\

 $c/a$ & 0.745 (0.296) & -3.69$\times$10$^{-2}$ (3.69$\times$10$^{-2}$) & 0.639 & 4.96$\times$10$^{-2}$ & 0.852\%\\

 \rowcolor{purple!5} $\rm log_{\rm 10}(\lambda_{\rm B})$ & 0.895 (0.349) & 6.13$\times$10$^{-2}$ (5.69$\times$10$^{-2}$) & 0.475 & 0.273 & 3.26\%\\

\end{tabular}
\caption{Same as \autoref{table:mwm_mass_frac_mod} but for cluster-mass haloes. These differences in medians are substantially smaller than the ones given in \autoref{subfigure:rhap_hist}.}
\label{table:rhap_mass_frac_mod}
\end{table*}

\section{Conclusion}
\label{Section:conclusion}
\subsection{Summary}
\label{subsection:summary}

In this extension to \citetalias{fielder2020}, we explore the influence of subhaloes on the properties of host dark matter haloes. In particular, we quantify how subhaloes affect three host halo properties: spin ($\lambda_{\rm B}$), concentration ($c_{\rm{NFW}}$), and shape, defined as the ratio of the shortest to longest halo axis (c/a). Throughout this paper, we refer to these three properties as our \textit{primary properties}. We calculate the value of each property for haloes in which both the smooth component belonging to the host, and subhaloes are included, and then recalculate them for \textit{only} the smooth component, this time excluding the mass belonging to subhaloes. We were able to separate subhaloes from their hosts by working directly with particle data from the Symphony simulations: the Milky Way-mass zoom-ins of \citet{mao2015} and the RHAPSODY cluster-mass simulations of \citet{wu2013}.  For a more in depth discussion with visuals of how halo density profiles are changed with the exclusion of subhaloes, we point the reader to \citetalias{fielder2020}.

For Milky Way mass haloes we found that upon removing subhaloes the median spin was reduced by $\sim$11\% concentration was increased by $\sim$30\%, and shape remained approximately unchanged. The cluster mass haloes saw similar trends with spin reduced by $\sim$27\%, concentration increased by $\sim$77\%, and shape increased by $\sim$11\%. We expand on these results by determining how these primary properties depend on other characteristics of the halo which we refer to as \textit{secondary properties}.

The secondary properties explored in this work are the fraction of halo mass contained in subhaloes and the halo mass accretion history. We find that there is a significant positive correlation between the fractional change of each of our primary properties and the mass fraction in subhaloes for both Milky Way- and cluster-mass haloes. The results for mass accretion history contain a lot of scatter and vary between the mass thresholds considered in this work and therefore we cannot draw any clear conclusions. 

We provide a means of correcting for the bias introduced by subhaloes in halo properties that were calculated including subhaloes. This is done by performing a linear regression with Huber's T estimator on the fractional change in primary properties as a function of the mass fraction in subhaloes. The parameters from this fit are used to form an equation to calculate corrections per halo. We apply these modifications on our own results and calculate RMSE to assess their performance. We find that RMSE values are small except in the case of concentration where there is some scatter.

\subsection{Discussion}
\label{subsection:discussion}
The conclusions discussed above hold consequences for dark matter halo studies. In particular, we show that haloes become more concentrated and that there is less scatter in concentration at fixed mass when subhaloes are removed. We interpret this as a result of the profile becoming more centrally peaked when mass located in the outer regions of the halo are removed, in agreement with findings in \citetalias{fielder2020}. Furthermore, we show that the change in concentration is a function of the fraction of mass removed, which is consistent with predictions from studies of the inverse relationship between halo mass and concentration. Some of the ongoing research on the concentration-mass relation has resulted in fitting relations between mass and concentration \citep[]{ludlow2014, Correa2015, Child2018}. However, these relations do not consider only the smooth component of the halo but substructure as well. Additionally, we detect a lot of scatter in the relationship between concentration change and mass assembly history which fits in with results from \citet{wang2020}.  

The results in this paper are relevant for observational and theoretical studies involving galaxies. The influence of subhaloes on host halo shape is especially important in studies involving elliptical galaxies. As discussed in \autoref{section:intro}, elliptical galaxies form as a product of merging galaxies originating from correlated directions and, as a result, resultant elliptical galaxies will have their longest axes point along the merger directions. We expect the same to hold true for the host halo, which we see in our results for cluster-mass haloes -- subhaloes induce ellipticity in their hosts. This bias in shape due to substructure is important in weak lensing analyses, where induced ellipticity due to mergers in coherent directions causes shear. Milky Way-mass haloes show little to no change in shape with the removal of subhaloes which indicates that subhaloes follow the smooth component ellipticity. 

For spiral galaxies, spin is a key defining property as in standard evolution theory the galaxy's angular momentum is predominantly set by that of the host halo in which it forms \citep{fall1980, 1997Dalcanton, mmw1998, Somerville_2008, dutton2012, 2013kravtsov}. For this reason, galaxy formation models use host halo spin as a proxy for galaxy spin. In this work, we show that for both Milky Way mass and cluster mass regimes, $\lambda_{\rm B}$ decreases as a result of removing subhaloes and the angular momentum they impart on the halo system. This indicates that subhaloes likely have greater velocities compared to their hosts. Therefore, we advise that disk galaxy spin should instead be inferred from the angular momentum of the smooth component of haloes only as subhaloes may bias spin.

The equations we provide for modifying halo data to exclude subhaloes can be of use in the research areas discussed above, in addition to other studies where a static halo potential, such as NFW, is assumed. One area where this assumption is made is in studies of stellar streams \citep{hendel2015, Sanderson2017, Bonaca2018, Dai2018}.

In summary, our results support those of \citetalias{fielder2020} in showing that halo concentration that can be significantly changed when subhalo matter is not included and additionally show that shape and spin are also affected. We likewise caution against including all substructure in properties calculations as it may result in biases in simulation and models as well as inconsistencies between that and observation. To this end, we provide modifications that can be applied to halo property calculations that were
made with the inclusion of subhaloes.

\section*{Acknowledgements}

The authors thank Oliver Hahn and Risa Wechsler for their 
contributions to the MMMZ and RHAPSODY simulations used in 
this research. The MMMZ and RHAPSODY simulations were created with computational resources at SLAC National Accelerator Laboratory, a U.S.\ Department of Energy Office; Y.-Y.M. and H.-Y.W.\ thank the support of the SLAC computational team. L.M. and A.R.Z. acknowledge support from the 
Department of Physics and Astronomy at the University of Pittsburgh and the 
Pittsburgh Particle Physics, Astrophysics, and Cosmology Center (Pitt PACC). 
K.W. acknowledges support from the Leinweber Foundation at the University of Michigan.
H.-Y.W.is supported by the DOE award DE-SC0021916.
This research made use of Python, along with many community-developed or maintained software packages, including
IPython \citep{ipython},
Jupyter (\http{jupyter.org}),
Matplotlib \citep{matplotlib},
NumPy \citep{numpy},
Pandas \citep{pandas},
and SciPy \citep{scipy}.
This research made use of NASA's Astrophysics Data System for bibliographic information.

\section*{Data Availability}
The Symphony data products used in this work are publicly available at \https{phil-mansfield.github.io/symphony}.

\bibliographystyle{mnras}
\bibliography{main}

\appendix
\section{Property Correlations with Mass Accretion History}
\label{sec:appendixA}

In this section we present \autoref{table:mwm_assembly} and \autoref{table:rhap_assembly}, which show the Spearman correlation test results for halo mass accretion history as discussed in \autoref{Section:imp_sec_prop} and \autoref{subsection:res_sec_prop}.

\begin{table*}
\centering
\large
\setlength{\tabcolsep}{22pt}
\renewcommand{\arraystretch}{1.5}
    \begin{tabular}{ c c c c}
 
    \multicolumn{4}{c}{\textsc{MMMZ Halo Property Correlation with Accretion History}} \\
    \rowcolor{purple!25} \textsc{Mass Acquired} & \textsc{Property} & \textsc{Coefficient} & \textsc{$p$-value}\\

    \cellcolor{purple!25}& $c_{\rm NFW}$ & 8.31$\times$10$^{-2}$ & 0.587 \\
    \cellcolor{purple!25}25\%  & c/a & 2.88$\times$10$^{-2}$ & 0.851 \\
    \cellcolor{purple!25}& $\rm log_{\rm 10}(\lambda_{\rm B})$ & 7.74$\times$10$^{-2}$ & 0.613 \\
    
    \rowcolor{purple!5}\cellcolor{purple!25}& $c_{\rm NFW}$ & -2.16$\times$10$^{-2}$ & 0.888 \\
    \rowcolor{purple!5}\cellcolor{purple!25}50\%  &c/a & 6.54$\times$10$^{-2}$ & 0.669\\
    \rowcolor{purple!5}\cellcolor{purple!25} & $\rm log_{\rm 10}(\lambda_{\rm B})$ & 0.233 &  1.23 \\

    \cellcolor{purple!25}& $c_{\rm NFW}$ & \textbf{0.312} & \textbf{3.71$\times$10$^{-2}$} \\
    \cellcolor{purple!25}70\%  & c/a & 0.254 & 9.27$\times$10$^{-2}$\\
    \cellcolor{purple!25}& $\rm log_{\rm 10}(\lambda_{\rm B})$ & \textbf{0.331} & \textbf{2.61$\times$10$^{-2}$}\\
  
    \rowcolor{purple!5}\cellcolor{purple!25} &  $c_{\rm NFW}$ & 0.337 & 2.36$\times$10$^{-2}$ \\
    \rowcolor{purple!5}\cellcolor{purple!25} 90\%  & c/a & 3.31$\times$10$^{-2}$ & 0.829 \\
    \rowcolor{purple!5}\cellcolor{purple!25} & $\rm log_{\rm 10}(\lambda_{\rm B})$ & 0.247 & 0.101\\
    
    \end{tabular}
\caption{Spearman correlation coefficients for fractional property change as a function of formation time, defined as the scale factor at which a halo acquires either 25\%, 50\%, 70\%, or 90\% of its mass, for Milky Way-mass haloes. Fractional change is defined as the residual fiducial host halo property relative to the property calculated without subhaloes. There is a significant positive relationship between concentration and formation time and spin and formation time defined only at the 70\% mass threshold.}
\label{table:mwm_assembly}
\end{table*}

\begin{table*}
\centering
\large
\setlength{\tabcolsep}{22pt}
\renewcommand{\arraystretch}{1.5}
\begin{tabular}{ c c c c}
 \multicolumn{4}{c}{\textsc{RHAPSODY Halo Property Correlation with Accretion History}} \\
 
 \rowcolor{purple!25}\textsc{Mass Acquired} & \textsc{Property} & \textsc{Coefficient} & \textsc{$p$-value}\\ 

    \cellcolor{purple!25}& $c_{\rm NFW}$ & 0.152 & 0.148\\
    \cellcolor{purple!25}25\%  & c/a & 0.101 & 0.339 \\
    \cellcolor{purple!25}& $\rm log_{\rm 10}(\lambda_{\rm B})$ & 0.178 & 0.089\\

    \rowcolor{purple!5}\cellcolor{purple!25} & $c_{\rm NFW}$ & 0.111 & 0.291\\
    \rowcolor{purple!5}\cellcolor{purple!25} 50\% & c/a & 6.03$\times$10$^{-2}$ & 0.568 \\
    \rowcolor{purple!5}\cellcolor{purple!25} &   $\rm log_{\rm 10}(\lambda_{\rm B})$ & 0.149 & 0.158\\

    \cellcolor{purple!25}&  $c_{\rm NFW}$ & 8.49$\times$10$^{-2}$ & 0.421 \\
    \cellcolor{purple!25}70\%  & c/a &  0.131 & 0.212 \\
    \cellcolor{purple!25}& $\rm log_{\rm 10}(\lambda_{\rm B})$ & 0.182 & 0.082\\

    \rowcolor{purple!5}\cellcolor{purple!25}&  $c_{\rm NFW}$ & -5.31$\times$10$^{-2}$ & 0.615 \\
    \rowcolor{purple!5}\cellcolor{purple!25}90\%  & c/a & 0.167 & 0.111 \\
    \rowcolor{purple!5}\cellcolor{purple!25}&  $\rm log_{\rm 10}(\lambda_{\rm B})$ & 0.181 & 0.085\\
    
\end{tabular}
\caption{Same as \autoref{table:mwm_assembly}, but for RHAPSODY haloes. There is no significant positive relationship between the three halo properties and formation time defined at any of the thresholds.}
\label{table:rhap_assembly}
\end{table*}

\section{Linear dependence of halo primary properties on subhalo mass fraction}
\label{sec:appendixB}

\begin{figure*}
\centering
\begin{subfigure}{0.48\textwidth}
    \includegraphics[width=\textwidth]{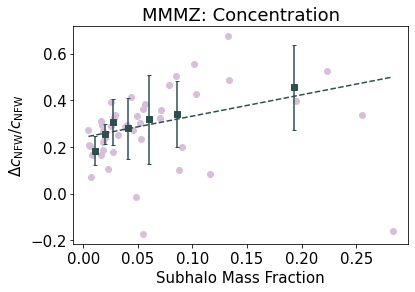}
    \caption{}
    \label{subfigure:mmmz_c_vs_mass_frac}
\end{subfigure}
\begin{subfigure}{0.48\textwidth}
    \includegraphics[width=\textwidth]{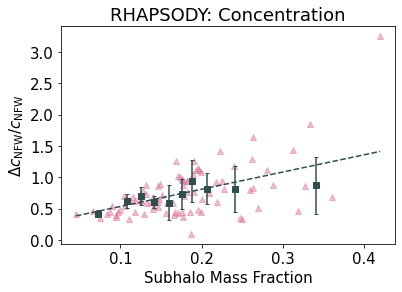}
    \caption{}
    \label{subfigure:rhap_c_vs_mass_frac}
\end{subfigure}
\begin{subfigure}{0.48\textwidth}
    \includegraphics[width=\textwidth]{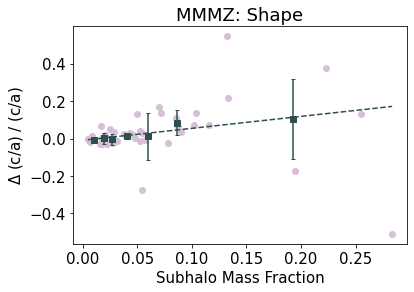}
    \caption{}
    \label{subfigure:mmmz_shape_vs_mass_frac}
\end{subfigure}
\begin{subfigure}{0.48\textwidth}
    \includegraphics[width=\textwidth]{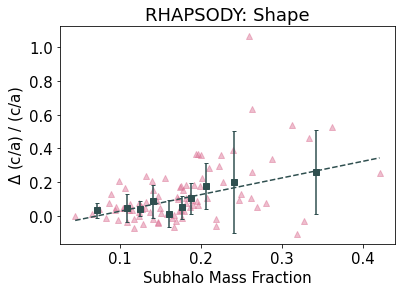}
    \caption{}
\label{subfigure:rhap_shape_vs_mass_frac}
\end{subfigure}
\begin{subfigure}{0.48\textwidth}
    \includegraphics[width=\textwidth]{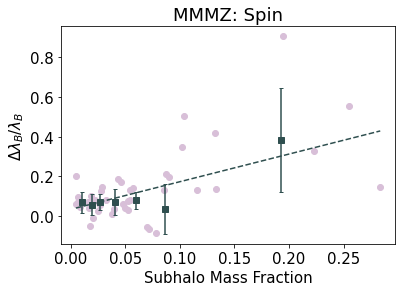}
    \caption{}
    \label{subfigure:mmmz_spin_vs_mass_frac}
\end{subfigure}
\begin{subfigure}{0.48\textwidth}
    \includegraphics[width=\textwidth]{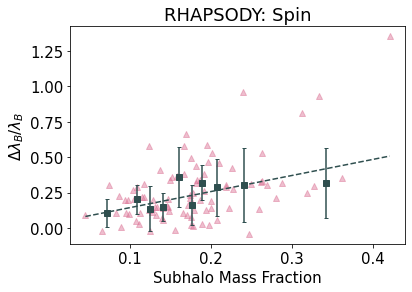}
    \caption{}
\label{subfigure:rhap_spin_vs_mass_frac}
\end{subfigure}
\caption{Scatter plots of the fractional change in concentration, shape, and spin as a function of subhalo mass fraction. Plots on the left correspond to the fit parameters given in \autoref{table:mwm_mass_frac_mod} and the plots on the right to those in \autoref{table:rhap_mass_frac_mod}. The purple circles (MMMZ) and pink triangles (RHAPSODY) represent individual haloes. Individual haloes are sorted into mass fraction bins with the dark gray squares showing the median value per bin and their standard deviation. The dark gray dashed line is the fit to the medians from the linear regression performed with Huber T's estimator.}
\label{figure:scatter_plots}
\end{figure*}

In this section we present \autoref{figure:scatter_plots}, which shows the linear regressions discussed in \autoref{subsubsection:corrections}.


\section{Inclusion of outliers}
\label{sec:appendixC}

In this section we present versions of \autoref{table:mwm_mass_frac_mod} and \autoref{table:rhap_mass_frac_mod} that are calculated without binning of data and with the inclusion of all outlier haloes that were not included in the main analysis with the exception of the one RHAPSODY halo with mislabelled host. Overall, results are qualitatively the same as in the main analysis. Slope and intercept values change when outliers are included but remain consistent and within error margins with those excluding outliers. For MMMZ haloes, the $\Delta$Median values in the rightmost column are smaller magnitude for concentration but greater for shape and approximately equal for spin. For RHAPSODY, $\Delta$Median and RMSE values for shape and concentration remain approximately the same, but spin $\Delta$Median decreases in magnitude whereas its RMSE value increases. In all cases, our modifications to halo properties are effective at reproducing property values without subhaloes.

\begin{table*}
\centering
\large
\setlength{\tabcolsep}{15pt}
\renewcommand{\arraystretch}{1.25}
\begin{tabular}{ cccccc }

 \multicolumn{6}{c}{\textsc{MMMZ Halo Property Modification with Outliers and No Bins
}}\\
 
 \rowcolor{purple!25} \textsc{Property} & \textsc{Slope} &\textsc{Intercept}
 &\textsc{$R^{2}$} & \textsc{RMSE} & \textsc{$\Delta$Median}\\

 \rowcolor{purple!5} $c_{\rm NFW}$ & 0.914 (0.364) & 0.241 (2.97$\times$10$^{-2}$) & -5.09$\times$10$^{-2}$ & 1.53 & -1.15\%\\

 $c/a$ & 0.638 (0.184) & -9.60$\times$10$^{-3}$ (1.45$\times$10$^{-2}$) & -5.49$\times$10$^{-2}$ & 6.94$\times$10$^{-2}$ & 0.227\%\\

 \rowcolor{purple!5} $\rm log_{\rm 10}(\lambda_{\rm B})$ & 1.39 (0.257) & 3.44$\times$10$^{-2}$ (2.09$\times$10$^{-2}$) & 0.366 & 0.193 & 0.597 \%\\

\end{tabular}
\caption{Modification parameters for each halo property for MMMZ haloes calculated without binning data. The first column shows the halo property and the second and third columns give the slope and intercept from the Huber's T linear regressions with their errors in parenthesis. The fourth column gives the \textsc{$R^{2}$} of these fits. We use the results from the fits to correct property measurements calculated including subhaloes. The RMSE from comparing the modified property values to the true values excluding subhaloes are given in the fifth column. The sixth column shows the percent change in median values from the true and fit-modified distribution. These differences are substantially smaller than the ones given in \autoref{subfigure:mmmz_hist}.}
\label{table:mwm_mass_frac_mod_w_outliers}
\end{table*}

\begin{table*}
\centering
\large
\setlength{\tabcolsep}{15pt}
\renewcommand{\arraystretch}{1.25}
\begin{tabular}{ cccccc }

 \multicolumn{6}{c}{\textsc{RHAPSODY Halo Property Modification with Outliers and No Bins}}\\
 \rowcolor{purple!25} \textsc{Property} & \textsc{Slope} &\textsc{Intercept} & \textsc{$R^{2}$} & \textsc{RMSE} & \textsc{$\Delta$Median}\\

 \rowcolor{purple!5} $c_{\rm NFW}$ & 2.22 (0.405) & 0.334 (8.16$\times$10$^{-2}$) & 0.321 & 1.35 & 4.39\%\\

 $c/a$ & 0.852 (0.182) & -5.12$\times$10$^{-2}$ (3.52$\times$10$^{-2}$) & 0.195 & 5.09$\times$10$^{-2}$ & 1.11\%\\

 \rowcolor{purple!5} $\rm log_{\rm 10}(\lambda_{\rm B})$ & 1.02 (0.232) & 4.75$\times$10$^{-2}$ (4.77$\times$10$^{-2}$) & 0.230 & 0.634 & 2.28\%\\

\end{tabular}

\caption{Same as \autoref{table:mwm_mass_frac_mod_w_outliers} but for cluster-mass haloes. These differences in medians are substantially smaller than the ones given in \autoref{subfigure:rhap_hist}.}
\label{table:rhap_mass_frac_mod_w_outliers}
\end{table*}

\section{Comments on shape}
\label{sec:appendixD}

In this section we discuss some of the nuance regarding calculating halo shape in N-body simulations. In our analysis we adopted the same algorithm as that in $\tt{ROCKSTAR}$ \citep{behroozi2013} where the calculation of shape is an iterative process. At each iteration, the halo axis are recalculated and particles that fall outside of the ellipsoid that is defined by these axis are removed. This process is repeated for several iterations until the change in shape between iterations is less than some tolerance. The process of removing particles at each iteration reduces the influence of subhaloes. Because subhaloes preferentially occupy more elliptical orbits, this mechanism influences halo shapes to be more spherical. In our analysis, we found that shape changes little when subhaloes are removed. We attribute this in part to the fact that subhalo influence is downplayed by the mechanism discussed above. To check whether this is the case, we recalculated shape without removing any particles and find that haloes were more elliptical.

Despite this, we chose to proceed with our original method for finding shape. There are multiple approaches for calculating halo shape which may all produce slightly different results. For example, \citet{allgood2006} uses the same iterative method as \citet{Bullock2002}, and the latter uses a spherical window. \citet{Kasun2005} use a spherical window like \citet{Bullock2002} but calculate shape by diagonalising the inertia tensor once rather than iterating. Another common method for calculating shape is by determining isodensity shells as a function radius as in \citet{jing2002}. For a more in depth description, we recommend reading Section 6 of \citet{allgood2006} which compares shape calculations using these various methods in more detail.

\section{Recalculating host halo size}
\label{sec:appendixE}

\begin{figure*}
\centering
\begin{subfigure}{\textwidth}
    \includegraphics[width=\textwidth]{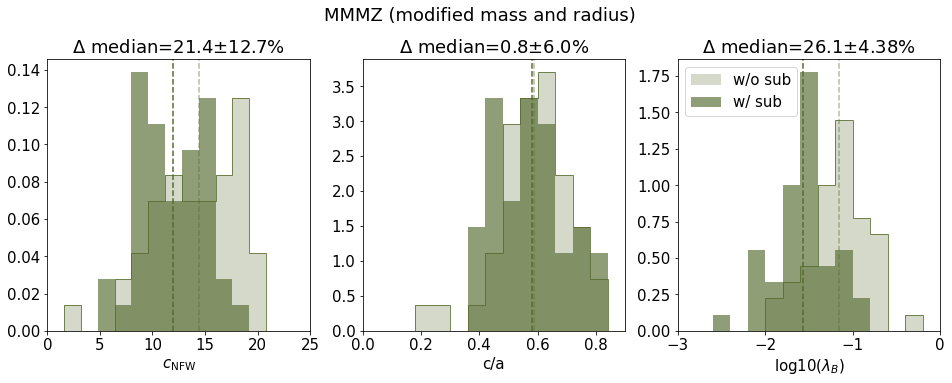}
    \caption{}
    \label{subfigure:mmmz_hist_new_rad}
\end{subfigure}
\begin{subfigure}{\textwidth}
    \includegraphics[width=\textwidth]{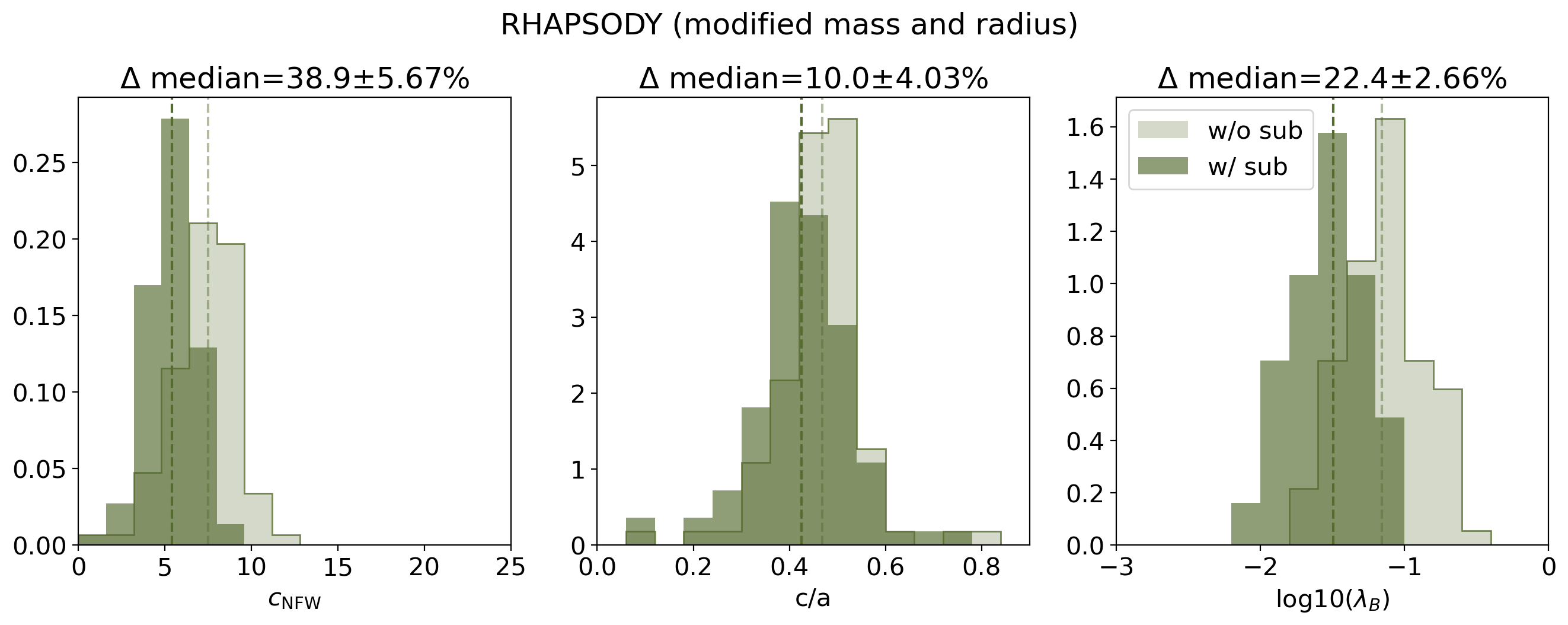}
    \caption{}
    \label{subfigure:rhap_hist_new_rad}
\end{subfigure}
\caption{Histograms of halo concentration, shape, and spin with and without the presence of subhaloes for the Milky Way-mass (a) and cluster mass (b) haloes \textit{using modified radii and masses for host haloes}. The $\Delta$ in the titles represents the difference in medians between the two distributions. The dark olive green histograms represent the halo properties including subhaloes (w/ sub) and the light olive green excluding subhaloes (w/o sub). The dashed vertical lines indicate the median value of each histogram. \textbf{(a):} In the left panel, concentration increases when subhaloes are not included as the density profile of the halo becomes more centrally dense. In the middle panel, there is no notable shift in the distribution of minor-to-major axis ratios. Finally, in the right panel, there is an increase in spin when subhaloes are removed. \textbf{(b)}: The lower row of panels depicts our results for the cluster-mass RHAPSODY haloes. In the left panel, concentration increases when subhaloes are not included as the density profile of the halo becomes more centrally dense. In the middle panel, there is a shift in the shape distribution towards unity, indicating that haloes are becoming more spherical. In the right panel, there is an increase in spin when subhaloes are removed.}
\label{figure:histograms_new_rvir}
\end{figure*}

\begin{figure*}
\centering
\begin{subfigure}{0.45\textwidth}
    \includegraphics[width=\textwidth]{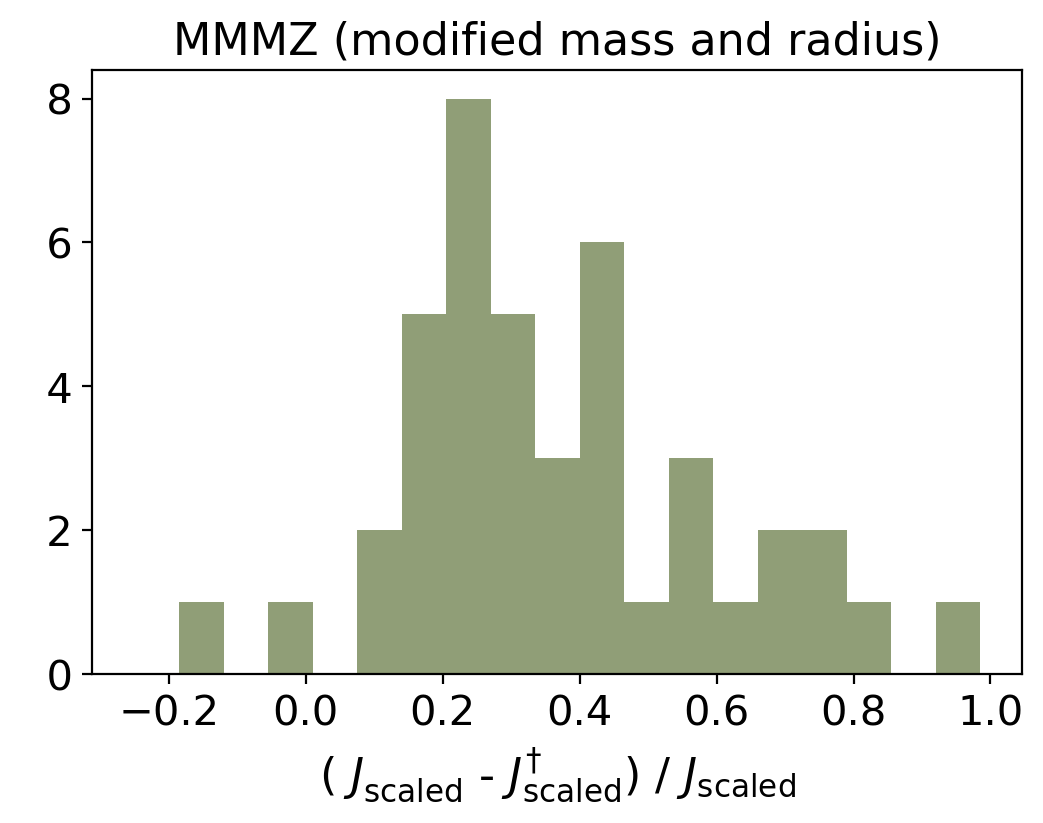}
    \caption{}
    \label{subfigure:mwm_ang_mom_diff_hist}
\end{subfigure}
\begin{subfigure}{0.45\textwidth}
    \includegraphics[width=\textwidth]{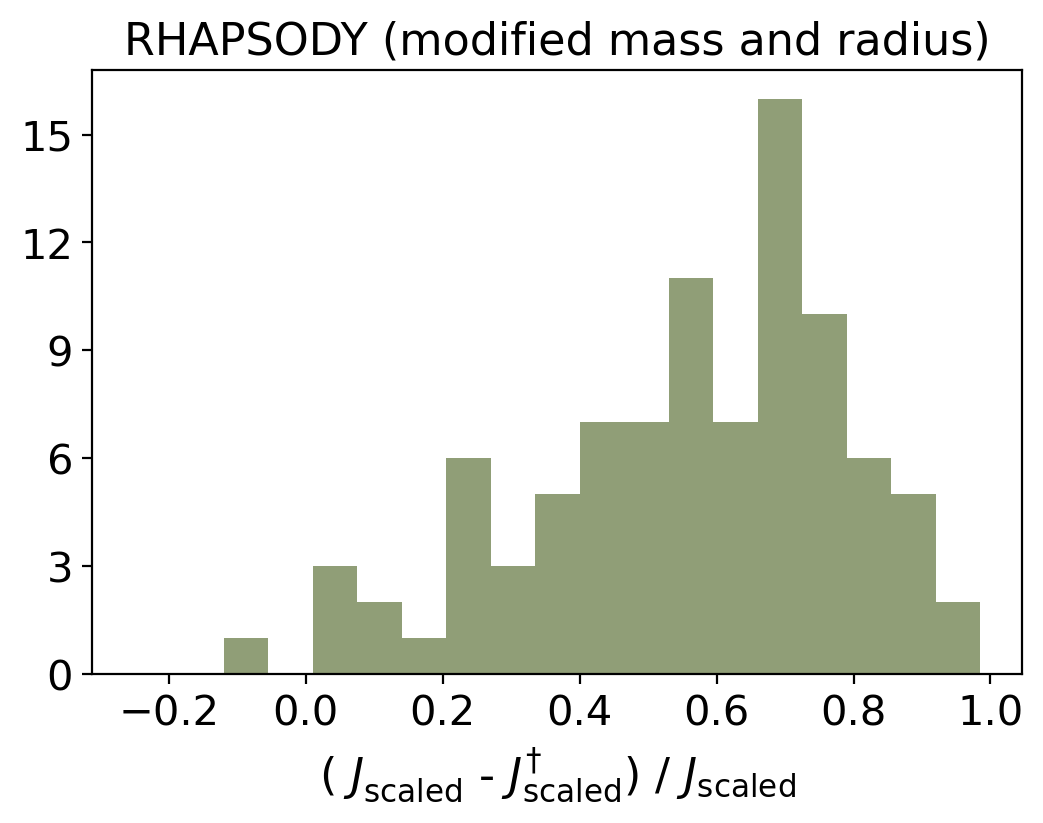}
    \caption{}
    \label{subfigure:rhap_ang_mom_diff_hist}
\end{subfigure}
\caption{Histograms of difference between angular momentum ($J$) including and excluding subhaloes \textit{relative} to that including subhaloes. $J^{\dagger}$ is calculated \textit{using modified radii and masses for host haloes}. The fiducial radius is 0.2Mpc for MMMZ and 1.7Mpc for RHAPSODY. For both MMMZ and RHAPSODY, the difference $J$-$J^{\dagger}$ is positive for all but a handful of haloes. This indicates that haloes generally lose angular momentum with the removal of subhaloes.}
\label{figure:ang_mom_histograms_new_rvir}
\end{figure*}

In the main text of this paper, we keep halo virial radius ($R_{\rm VIR}$) and virial mass ($M_{\rm VIR}$) fixed throughout all calculations despite removing mass contained in subhalos. In this appendix we repeat the analysis in \autoref{subsection:imp_primary}, however, this time host properties are calculated using updated virial radii and masses that depend on particles associated with the host halo only using the {\rm subhalo excluded} set of particles. The new virial radius is defined as the radius at which the mean overdensity of {\em subhalo excluded} particles enclosed by a sphere centered on the host is equal to $\Delta_{\rm VIR}$$\rho_{\rm VIR}$. Given the cosmological parameters of the RHAPSODY and MMMZ simulations, $\Delta_{\rm VIR}$ = 94. The new virial mass is defined as the number of particles enclosed by $R_{\rm VIR}$ multiplied by the mass of each particle. In recalculating $R_{\rm VIR}$, we find that it is $\sim$4\% smaller for MMMZ haloes, and $\sim$13\% smaller for RHAPSODY. Results are shown in \autoref{figure:histograms_new_rvir} and \autoref{figure:ang_mom_histograms_new_rvir}.

In \autoref{figure:histograms_new_rvir} we see that changes in concentration and shape remain qualitatively unchanged from the main text (see \autoref{figure:histograms}, however, the magnitude of the change is altered for concentration. For RHAPSODY, concentration now increases by $\sim$38\% rather than $\sim$70\%. Change in shape remains at approximately 10\% and is consistent within error margins regardless of how host radius and mass is defined. The property that is most altered by using the new radii and masses is spin; rather than {\em decreasing} by $\sim$26\% when subhaloes are removed, it now {\em increases} by $\sim$22\%. We attribute this to the way in which angular momentum is normalised in the equation for spin (see \autoref{eq:spin}).
According to the definition of virial mass and virial radius, 
$R_{\mathrm{VIR}} \propto M_{\mathrm{VIR}}^{1/3}$ and $V_{\mathrm{VIR}} \propto M_{\mathrm{VIR}}^{1/3}$, so that the denominator in \autoref{eq:spin} scales with 
mass as $M_{\mathrm{VIR}}V_{\mathrm{VIR}}R_{\mathrm{VIR}} \propto M_{\mathrm{VIR}}^{5/3}$. 
This sensitivity of the normalization of the spin parameter to the host halo mass means that, 
in practice, the decrease in mass due to the removal of subhalo particles reduces the 
normalisation of the spin parameter significantly more than the removal of subhaloes 
reduces the angular momentum.
This is illustrated in \autoref{subfigure:rhap_ang_mom_hist}, 
which depicts the general decline in angular momentum upon removal of 
subhaloes. We expand on this in \autoref{subfigure:rhap_ang_mom_diff_hist}, 
where we show that the angular momentum, $J$, 
declines when subhaloes are removed for the vast majority of host haloes.

The host haloes of the MMMZ sample behave similarly. Concentration increases by $\sim$21\% 
rather than $\sim$30\% and shape remains unchanged. The spin of the MMMZ haloes now {\em increases}
by $\sim$26\% when subhaloes are removed rather than {\em decreasing} by $\sim$10\%. In \autoref{subfigure:mwm_ang_mom_diff_hist} we show that the difference in $J$ with and without subhaloes is positive for all but a few haloes. Host haloes generally lose angular 
momentum when subhaloes are removed.

\bsp	
\label{lastpage}
\end{document}